\documentclass[manuscript,onefignum,onetabnum]{siamart190516}

\usepackage[a4paper, total={6in, 8in}]{geometry}
\usepackage{graphicx}
\usepackage{subfig}
\usepackage{amssymb}
\usepackage{xcolor}
\usepackage{soul}
\usepackage{amsmath}
\usepackage{lineno}
\usepackage{graphics}
\usepackage{caption}
\usepackage{lscape}
\usepackage{tabularx}
\usepackage[makeroom]{cancel}
\usepackage[english]{babel}
\usepackage{float}
\usepackage{amsfonts}
\usepackage{textcomp}
\usepackage[utf8]{inputenc}
\usepackage[english]{babel}
\usepackage{mathtools}
\usepackage{lipsum}
\usepackage{todonotes}

\usepackage{algorithm}
\usepackage{algorithmic}
\usepackage{booktabs}

\usepackage{lipsum}
\usepackage{amsfonts}
\usepackage{graphicx}
\usepackage{epstopdf}

\ifpdf
  \DeclareGraphicsExtensions{.eps,.pdf,.png,.jpg}
\else
  \DeclareGraphicsExtensions{.eps}
\fi

\newsiamremark{remark}{Remark}
\newsiamremark{hypothesis}{Hypothesis}
\crefname{hypothesis}{Hypothesis}{Hypotheses}
\newsiamthm{claim}{Claim}

\headers{Grassmannian diffusion maps}{dos Santos, Giovanis, Kontolati, Loukrezis, and Shields}

\title{Grassmannian diffusion maps based surrogate modeling via geometric harmonics
}

\author{Ketson R. M. dos Santos\footnotemark[1]
\and Dimitrios G. Giovanis\footnotemark[2]
\and Katiana Kontolati\footnotemark[2]
\and Dimitrios Loukrezis\footnotemark[3]\ \footnotemark[4]
\and Michael D. Shields\footnotemark[2]
}

\usepackage{amsopn}

\ifpdf
\hypersetup{
  pdftitle={Grassmannian diffusion maps based surrogate modeling via geometric harmonics},
  pdfauthor={K. R. M. dos Santos, D. G. Giovanis, K. Kontolati, D. Loukrezis, and M. D. Shields}
}
\fi

\begin{document}

\maketitle

\begin{abstract}
{In this paper, a novel surrogate model based on the Grassmannian diffusion maps (GDMaps) and utilizing geometric harmonics is developed for predicting the response of engineering systems and complex physical phenomena. The method utilizes the GDMaps to obtain a low-dimensional representation of the underlying behavior of physical/mathematical systems with respect to uncertainties in the input parameters. Using this representation, geometric harmonics, an out-of-sample function extension technique, is employed to create a global map from the space of input parameters to a Grassmannian diffusion manifold. Geometric harmonics is also employed to locally map points on the diffusion manifold onto the tangent space of a Grassmann manifold. The exponential map is then used to project the points in the tangent space onto the Grassmann manifold, where reconstruction of the full solution is performed. The performance of the proposed surrogate modeling is verified with three examples. The first problem is a toy example used to illustrate the development of the technique. In the second example, errors associated with the various mappings employed in the technique are assessed by studying response predictions of the electric potential of a dielectric cylinder in a homogeneous electric field. The last example applies the method for uncertainty prediction in the strain field evolution in a model amorphous material using the shear transformation zone (STZ) theory of plasticity. In all examples, accurate predictions are obtained, showing that the present technique is a strong candidate for the application of uncertainty quantification in large-scale models.}

\end{abstract}

\begin{keywords}
  Surrogate modeling, manifold learning, diffusion maps, Grassmann manifold, geometric harmonics
\end{keywords}


\section{Introduction}
\renewcommand{\thefootnote}{\fnsymbol{footnote}}
\footnotetext[1]{Earthquake Engineering and Structural Dynamics Laboratory, \'{E}cole Polytechnique F\'{e}d\'{e}rale de Lausanne, VD, Switzerland.}
\footnotetext[2]{Department of Civil \& Systems Engineering, Johns Hopkins University, Baltimore, MD, 21218, USA.}
\footnotetext[3]{Institute for Accelerator Science and Electromagnetic Fields (TEMF), Technische Universität Darmstadt, Darmstadt, Germany.}
\footnotetext[4]{Centre for Computational Engineering, Technische Universität Darmstadt, Darmstadt, Germany}
\label{S:Intro}
Surrogate models (aka emulators or metamodels) have become an important tool for uncertainty quantification because they afford a computationally efficient means of approximating (often complex) input-output relations generated from high-fidelity computational models. Surrogate models are constructed by training (or learning) a mathematical model from a finite set of input-output observations referred to as the training set. Given a new set of input parameters, where the solution of the model is unknown, the surrogate can then be used to predict the solution at minimal cost. Surrogate models are typically classified as either intrusive or non-intrusive \cite{lemaitre2010,smith2013}. Intrusive methods, such as those based on Galerkin schemes, typically provide good convergence \cite{ghanem2003, babuska2010}, but their complexity limits their flexibility \cite{chen2015} and requires the development of new numerical schemes and, hence, entirely new solvers. In contrast, non-intrusive methods are trained from realizations of a deterministic solver (i.e. they can leverage pre-existing numerical methods and solvers) at selected sample points generated from the uncertain parameters. Moreover, surrogate models often provide an improvement over statistical approaches based on Monte Carlo simulation (MCS). Although MCS is a versatile and non-intrusive method, it typically offers slow convergence with the number of samples \cite{fishman1996}. To circumvent this limitation, quasi Monte Carlo methods \cite{caflisch1998}, adaptive sampling techniques \cite{shields2015,shields2018}, and other intelligent sampling techniques leveraging variance reduction techniques \cite{mckay1979,echard2011,santos2015,botev2017} are employed. Increasingly, these enhanced sampling methods are being leveraged to improve the training efficiency of surrogate models; enabling surrogates to be developed from far fewer training data.
 
Among the most widely-used surrogate models in UQ are polynomial chaos expansions (PCE) \cite{ghanem2003} and Gaussian process (GP) regression (Kriging) \cite{krige1951,rasmussen2004}. PCE was originally proposed by Wiener \cite{wiener1938} based on the projection of random solutions onto a basis of Hermite polynomials, which are orthogonal with respect to the Gaussian measure. Ghanem et al. \cite{ghanem2003} introduced the Stochastic Galerkin projection for PCE, an intrusive method that requires the formulation of a system of algebraic equations for distinct classes of problems. The generalized PCE (gPCE) \cite{xiu2003} provides improved convergence and enhanced flexibility by utilizing polynomial bases from the Wiener-Askey scheme. Moreover, gPCE can be used in the construction of non-intrusive surrogate based on collocation schemes aiming at enhanced versatility and convergence properties. On the other hand, a GP approximates the input-output relation as a Gaussian stochastic process completely described by its mean and correlation function. Learning a GP thus consists of determining the mean and correlation functions. However, the computational complexity of the GP training process with $n$ observations is on the order $\mathcal{O}(n^3)$, whereas the memory requirements scale with $\mathcal{O}(n^2)$.

It is well known that the relationship between data dimensionality and model fidelity has a strong influence on the computational performance of UQ for large-scale models. In this regard, dimension reduction techniques have become attractive due to their ability to represent high-dimensional data in a low-dimensional and more informative space (manifold) \cite{nadler2006,erban2007,coifman2008,dsilva2015}. Dimension reduction techniques can be classified into linear and nonlinear methods. Linear methods include principal component analysis (PCA) \cite{pearson1901}, locality preserving projections (LPP) \cite{he2005}, linear regression \cite{rangarajan2005}, and singular value decomposition \cite{strang2016}. On the other hand, nonlinear methods are useful for constructing nonlinear maps between the high- and low-dimensional spaces. Nonlinear methods include Isomaps \cite{tenenbaum2000}, locally linear embedding (LLE) \cite{roweis2000,donoho2003}, Kernel PCA \cite{lataniotis2020}, and diffusion maps (DMaps) \cite{coifman2005,coifman2006}. It is worth noting that several methods based on DMaps have been proposed in the literature such as the method proposed by Soize and Ghanem \cite{soize2016sampling} for sampling a random vector whose probability distribution is constrained to an Euclidean manifold. They used DMaps for discovering the underlying structure of the dataset. Moreover, DMaps has been used in the development of surrogate models either based on local polynomial interpolations and the Nystr\"{o}m out-of-sample extension \cite{kalogeris2020}, or based on neural networks and the Laplacian pyramids \cite{kalogeris2021}. Further, a subspace extension of Diffusion Maps, the Grassmannian Diffusion Maps (GDMaps), was introduced by dos Santos et al. \cite{santos2020}. In this technique, a model dimension hyper-reduction is achieved by combining a pointwise linear dimensionality reduction technique, projecting the high-dimensional data onto a low-dimensional Grassmann manifold, and a multipoint nonlinear dimension reduction using Diffusion maps which reveals the intrinsic structure of the data on the manifold. Recently, Kontolati et al. \cite{kontolati2021} employed GDMaps and PCE in the development of a manifold learning based method for UQ in systems describing complex processes.

In this work, we leverage the GDMaps to construct surrogate models for very high-dimensional systems. More specifically, a set of low-dimensional coordinates embedding the high-dimensional model data on the Grassmann manifold is obtained via Grassmannian diffusion maps (GDMaps). This low-dimensional diffusion space serves as a connecting space between the parameter space and the Grassmann manifold, where full solution reconstruction can be performed. To connect these spaces, maps are constructed using the idea of out-sampling extension \cite{burt1983, broomhead1988,mahdisova2017,leeb2019}. Inspired by the the Nystr\"{o}m extension \cite{williams2001Nystrom,heimowitz2018}, Coifman and Lafon \cite{coifman2006gh} introduced a scheme, referred to as geometric harmonics (GH) for extending empirical functions only available at few locations. They demonstrated that this process relates the function complexity and its extension, with important implications to the construction of the lifting and restriction operators. In this paper, GH is employed in the construction of an out-of-sample extension to create maps between spaces of interest. A global GH surrogate is constructed between the parameter and diffusion spaces, while local GH surrogates are constructed between the diffusion space and the Grassmann manifold via the tangent space, a flat inner-product space allowing local exponential mapping onto the Grassmannian, where the predicted model response can be constructed. Moreover, GH is also used to create a map between the parameter space and the space of singular values used in the contruction of the model prediction.

This paper is organized as follows. Section 2 discusses important background on the Grassmann manifold. Section 3 provides an overview of the Grassmannian diffusion maps technique. Section 4 introduces GH as an out-of-sample function extension technique. Section 5 contains a detailed description of the surrogate modeling approach, referred to as Grassmannian-Geometric Harmonics Maps (Grassmannian-GHMaps), developed herein. In section 6, three examples are provided. The first is a simple example used to explain the proposed method in a manner that is conceptually understandable and easy to visualize. The second example uses the GDMaps surrogates to predict the electric potential for a dielectric cylinder in homogeneous electric field, and is used to perform error analysis on the proposed method. The third example develops a surrogate model for the evolution of the strain field of amorphous solids under simple shear using the shear transformation zone (STZ) theory of plasticity. Finally, concluding remarks are provided in Section 7. Furthermore, the algorithms presented in this paper have been implemented in UQpy (Uncertainty Quantification with python) \cite{olivier2020uqpy}, a general purpose Python toolbox for modeling uncertainty in physical and mathematical systems.

\section{Grassmann manifold}
\label{s:grassmann}
The concepts presented in this section are essential for the development of a surrogate model based on the Grassmannian diffusion maps. Let us begin by defining a $p$-plane as a $p$-dimensional subspace, and a $p$-frame as a coordinate system that spans that subspace. Based on these two concepts one can define two important manifolds, the Stiefel and the Grassmann manifold. The Stiefel manifold $\mathcal{V}(p,n)$ is the set of all $p$-frames in $\mathbb{R}^n$ such that $\mathcal{V}(p,n) = \{\mathbf{X} \in \mathbb{R}^{n \times p}: \mathbf{X}^\intercal\mathbf{X} = \mathbf{I}_p\}$, where $\mathbf{I}_p \in \mathbb{R}^{p \times p}$ is the identity matrix and $\mathbf{X} \in \mathbb{R}^{n \times p}$ is an orthonormal matrix \cite{auslander2012}. The Grassmann manifold (or Grassmannian) $\mathcal{G}(p,n)$ is the set of $p$-planes in $\mathbb{R}^n$, where a point is given by $\mathcal{X} = \mathrm{span}\left(\mathbf{\Psi}\right)$, with $\mathbf{\Psi} \in \mathcal{V}(p,n)$ \cite{ye2014}. It is worth noting that $\mathcal{X}$ is identified as an equivalence class of $n \times p$ matrices under orthogonal transformation of the Stiefel manifold \cite{ye2014,ye2019,lek2019}. Therefore, a point on the Grassmann manifold is represented by an orthonormal matrix $\mathbf{\Psi} \in \mathbb{R}^{n \times p}$ (the Stiefel representation). 

\subsection{Tangent Space: Exponential and logarithmic maps}
\label{s:exp_log}
The Grassmann manifold is a smooth and continuously differentiable manifold, which enables numerous mathematical operations such as differentiation and optimization \cite{edelman1998geometry,amsallem2008interpolation}. Given that the Grassmann manifold is smooth and continuously differentiable, one can define a trajectory $\gamma(z), z\in[0,1]$, known as geodesic, defining the shortest path between two points, $\gamma(0)=\mathcal{X}_0 = \mathrm{span}(\mathbf{\Psi}_0)$ and $\gamma(1)=\mathcal{X}_1 = \mathrm{span}(\mathbf{\Psi}_1)$, on the manifold $\mathcal{G}(p,n)$ \cite{edelman1998geometry}. The derivative of this trajectory at any point $\mathcal{X}$ (represented by $\mathbf{\Psi}$) defines the tangent space ($\mathcal{T}_{\mathcal{X}} \mathcal{G}(p,n)$), which is given by the set of all tangent vectors $\mathbf{\Gamma}$ such that $\mathcal{T}_{\mathcal{X}} \mathcal{G}(p,n) = \{ \mathbf{\Gamma} \in \mathbb{R}^{n \times p}: \mathbf{\Gamma}^\top \mathbf{\Psi} = 0 \}$. Therefore, given a tangent space $\mathcal{T}_{\mathcal{X}_0}\mathcal{G}(p,n)$ at $\mathcal{X}_0$, one can map $\mathbf{\Gamma}_1$ onto the Grassmannian point $\gamma(1) = \mathcal{X}_1$ represented by $\mathbf{\Psi}_1$ via the exponential map

\begin{equation}\label{eq:2.7}
    \mathbf{\Psi}_1 = \mathrm{exp}_{\mathcal{X}_0}(\mathbf{\Gamma}_1)= \mathrm{exp}_{\mathcal{X}_0}(\mathbf{U}\mathbf{S}\mathbf{V}^T) = \mathbf{\Psi}_0\mathbf{V}\mathrm{cos}\left(\mathbf{S}\right)\mathbf{Q}^T+\mathbf{U}\mathrm{sin}\left(\mathbf{S}\right)\mathbf{Q}^T,
\end{equation}
where $\mathbf{Q} \in \mathbb{R}^{n \times n}$ is an orthogonal matrix satisfying the following expressions.
\begin{equation}\label{eq:2.8}
    \mathbf{V}\mathrm{cos}\left(\mathbf{S}\right)\mathbf{Q}^T = \mathbf{\Psi}_0^T\mathbf{\Psi}_1,
\end{equation}
\noindent
and
\begin{equation}\label{eq:2.9}
    \mathbf{U}\mathrm{sin}\left(\mathbf{S}\right)\mathbf{Q}^T = \mathbf{\Psi}_1 -  \mathbf{\Psi}_0\mathbf{\Psi}_0^T\mathbf{\Psi}_1.
\end{equation}
\noindent
After appropriate manipulation, one can obtain the following expression.
\begin{equation}\label{eq:2.10}
    \mathbf{U}\mathrm{tan}\left(\mathbf{S}\right)\mathbf{V}^T = \left(\mathbf{\Psi}_1 -  \mathbf{\Psi}_0\mathbf{\Psi}_0^T\mathbf{\Psi}_1\right)\left(\mathbf{\Psi}_0^T\mathbf{\Psi}_1\right)^{-1}.
\end{equation}
\noindent
Consequently, one can write the logarithmic map from the Grassmannian to the tangent space, $\mathrm{log}_\mathcal{X}:\mathcal{G}(p,n) \rightarrow \mathcal{T}_{\mathcal{X}}\mathcal{G}(p,n)$ as
\begin{equation}\label{eq:2.11}
    \mathrm{log}_\mathcal{X}(\mathbf{\Psi}_1) = \mathbf{U}\mathrm{tan}^{-1}\left(\mathbf{S}\right)\mathbf{V}^T.
\end{equation}

\subsection{Grassmannian distance}
\label{s:gr_dist}
The properties of the Grassmann manifold further afford a notion of distance between points on it. Many definitions of distance exist and can be expressed in terms of the principal angles between subspaces \cite{hamm2008}. One can easily see that the cosine of the principal angles $\theta_i \in \left[0, \pi/2\right]$ between two subspaces $\mathcal{X} = \mathrm{span}(\mathbf{\Psi}_x)$ and $\mathcal{Y} = \mathrm{span}(\mathbf{\Psi}_y)$ can be computed from the singular values of $\mathbf{\Psi}_x^T\mathbf{\Psi}_y = \mathbf{\bar{U}}\mathbf{\bar{S}}\mathbf{\bar{V}}^T$, where $\mathbf{\bar{U}} \in O(k)$, $\mathbf{\bar{V}} \in O(l)$, and $\mathbf{\bar{S}} = \mathrm{diag}(\sigma_1, \sigma_2, \dots, \sigma_p)$, with $p = \mathrm{min}(k,l)$. Thus, the principal angles are computed as $\theta_i = \mathrm{cos}^{-1}(\sigma_i)$ \cite{miao1992}. A well-known and commonly used distance is the geodesic distance, $d_{\mathcal{G}(p,n)}\left(\mathcal{X},\mathcal{Y}\right)$, corresponding to the distance along the geodesic curve $\gamma(z)$ parameterized by $z \in [0, 1]$, and given by $d_{\mathcal{G}(p,n)}\left(\mathcal{X},\mathcal{Y}\right) = \|\mathbf{s}\|_2$ \cite{wong1967,ye2014,giovanis2018}, where $\mathbf{s} = \left(\theta_1, \theta_2, \dots, \theta_p \right)$ is the vector of principal angles. See \cite{santos2020} for more definitions of distances on the Grassmann manifold.

\subsection{Karcher mean}
\label{s:gr_km}
Consider a set of points (subspaces) on $\mathcal{G}(p,n)$. The Riemannian center of mass of these points is known as Karcher mean, $\mu_{\mathcal{G}(p,n)}$, and 
corresponds to the point $\mathcal{Y} \in \mathcal{G}(p,n)$ that minimizes the cost function $\sigma^2_{\mathcal{G}(p,n)}: \mathcal{G}(p,n) \rightarrow \mathbb{R}^{+}$ \cite{karcher1977,giovanis2020} given by
\begin{equation}\label{eq:lambda_cf}
    \sigma^2_{\mathcal{G}(p,n)}(\mathcal{Y}) = \int_{\mathcal{G}(p,n)}d^2_{\mathcal{G}(p,n)}(\mathcal{Y},\mathcal{X})dP(\mathcal{X}),
\end{equation}
where $dP(\mathcal{X}) = \rho(\mathcal{X})d\mathcal{G}(p,n)$ is a probability measure on the Grassmann manifold with probability density function $\rho(\mathcal{X})$.  Thus, the Karcher mean $\mu_{\mathcal{G}(p,n)}=\mathrm{span}(\mathbf{M})$ can be computed by solving the optimization problem
\begin{equation}\label{eq:min_cf}
    \mu_{\mathcal{G}(p,n)} \doteq \underset{\mathcal{Y} \in \mathcal{G}(p,n)}{\mathrm{argmin}}\int_{\mathcal{G}(p,n)}d^2_{\mathcal{G}(p,n)}(\mathcal{Y},\mathcal{X})dP(\mathcal{X}).
\end{equation}

One can easily notice a similarity between Eq. (\ref{eq:lambda_cf}) and the variance of a continuous random variable, which gives a notion of dispersion around the mean. The Karcher variance $\sigma^2_{\mathcal{G}(p,n)}$ and the Karcher mean $\mu_{\mathcal{G}(p,n)}$ can therefore be interpreted as the mean and variance of a continuous random variable on $\mathcal{G}(p,n)$. With this interpretation, the Karcher mean can be estimated from a discrete set of points on the Grassmann manifold $S=\{\mathcal{X}_1, \dots, \mathcal{X}_N\} \subset \mathcal{G}(p,n)$, by solving the following minimization
\begin{equation}\label{eq:min_cf_disc}
    \hat{\mu}_{\mathcal{G}(p,n)} \approx \underset{\mathcal{Y} \in \mathcal{G}(p,n)}{\mathrm{argmin}}\frac{1}{N}\sum^N_{i=1}d^2_{\mathcal{G}(p,n)}(\mathcal{Y},\mathcal{X}_i).
\end{equation}

\subsection{Grassmannian kernels}
\label{S:gr_kernels}
Kernel-based dimensionality reduction techniques such as the conventional diffusion maps depend on the appropriate definition of a real-valued positive semi-definite kernel $k(x_i,x_j)$ with $\sum_{i,j}c_ic_jk(x_i,x_j) \leq 0$, where $c_i,c_j \in \mathbb{R}$. In this regard, the Gaussian kernel (Eq. \ref{eq:gaussian_kernel}) is perhaps the most widely used.
\begin{equation}\label{eq:gaussian_kernel}
    k(\mathbf{X}_i,\mathbf{X}_j) = \mathrm{exp}\left( -\frac{||\mathbf{X}_i - \mathbf{X}_j||^2_2}{4 \epsilon}\right),
\end{equation}
where $\mathbf{X}_i$ and $\mathbf{X}_j$ are the high-dimensional data and $\epsilon$ is a length-scale parameter. However, the Gaussian kernel is not suitable to represent the underlying subspace structure of datasets. On the other hand, Grassmannian kernels are endowed with this feature, which is advantageous in the analysis of high-dimensional data. A Grassmannian kernel is defined as a real symmetric map $k: \mathcal{G}(p,n) \times \mathcal{G}(p,n) \rightarrow \mathbb{R}$ embedding the Grassmann manifold into a reproducing kernel Hilbert space. Moreover, a Grassmannian kernel is invariant to the choice of basis and is positive semi-definite. Several families of Grassmannian kernels, with different characteristics, are proposed in the literature (see \cite{hamm2008, hamm2009, harandi2014}). However, the most popular kernels are the Binet-Cauchy and Projection kernels. The Binet-Cauchy kernel is constructed by embedding the Grassmann manifold $\mathcal{G}(p,n)$ into a projective space $\mathbb{P}\left(\bigwedge^{\raisebox{-0.4ex}{\scriptsize $p$}} \mathbb{R}^{n}\right)$. Considering two subspaces $\mathcal{X} = \mathrm{span}(\mathbf{\Psi}_x)$ and $\mathcal{Y} = \mathrm{span}(\mathbf{\Psi}_y)$, the Binet-Cauchy kernel is given by
\begin{equation}\label{eq:kernel_bc}
    k_{bc}(\mathcal{X},\mathcal{Y}) = \mathrm{det}\left(\mathbf{\Psi}_x^T\mathbf{\Psi}_y\right)^2,
\end{equation}
\noindent
or equivalently in terms of principal angles \cite{hamm2008,harandi2014}
\begin{equation}\label{eq:kernel_bc_angles}
    k_{bc}(\mathcal{X},\mathcal{Y}) = \prod_{i=1}^{p}\mathrm{cos}^2(\theta_i).
\end{equation}

Another kernel frequently used in kernel-based methods on the Grassmann manifold is the projection kernel. It is constructed based on the projection embedding $\Pi: \mathcal{G}(p,n) \rightarrow \mathbb{R}^{n \times n}$ such that $\Pi \left(\mathbf{\Psi}\right) = \mathbf{\Psi}\mathbf{\Psi}^T$. This kernel is defined as
\begin{equation}\label{eq:kernel_proj}
    k_{pr}(\mathcal{X},\mathcal{Y}) = ||\mathbf{\Psi}_x^T\mathbf{\Psi}_y||_F^2,
\end{equation}
\noindent
or equivalently in terms of principal angles \cite{hamm2008,harandi2014}
\begin{equation}\label{eq:kernel_proj_angles}
    k_{pr}(\mathcal{X},\mathcal{Y}) = \sum_{i=1}^{p}\mathrm{cos}^2(\theta_i).
\end{equation}

Further discussion of Grassmannian kernels and their specific use for Grassmannian diffusion maps can be found in dos Santos et al. \cite{dossantos2020}.

\section{Grassmannian diffusion maps}
\label{S:GDMAPs}
\begin{figure}[!ht]
	\centering
	\captionsetup{justification=centering}
	\includegraphics[width=0.5\textwidth]{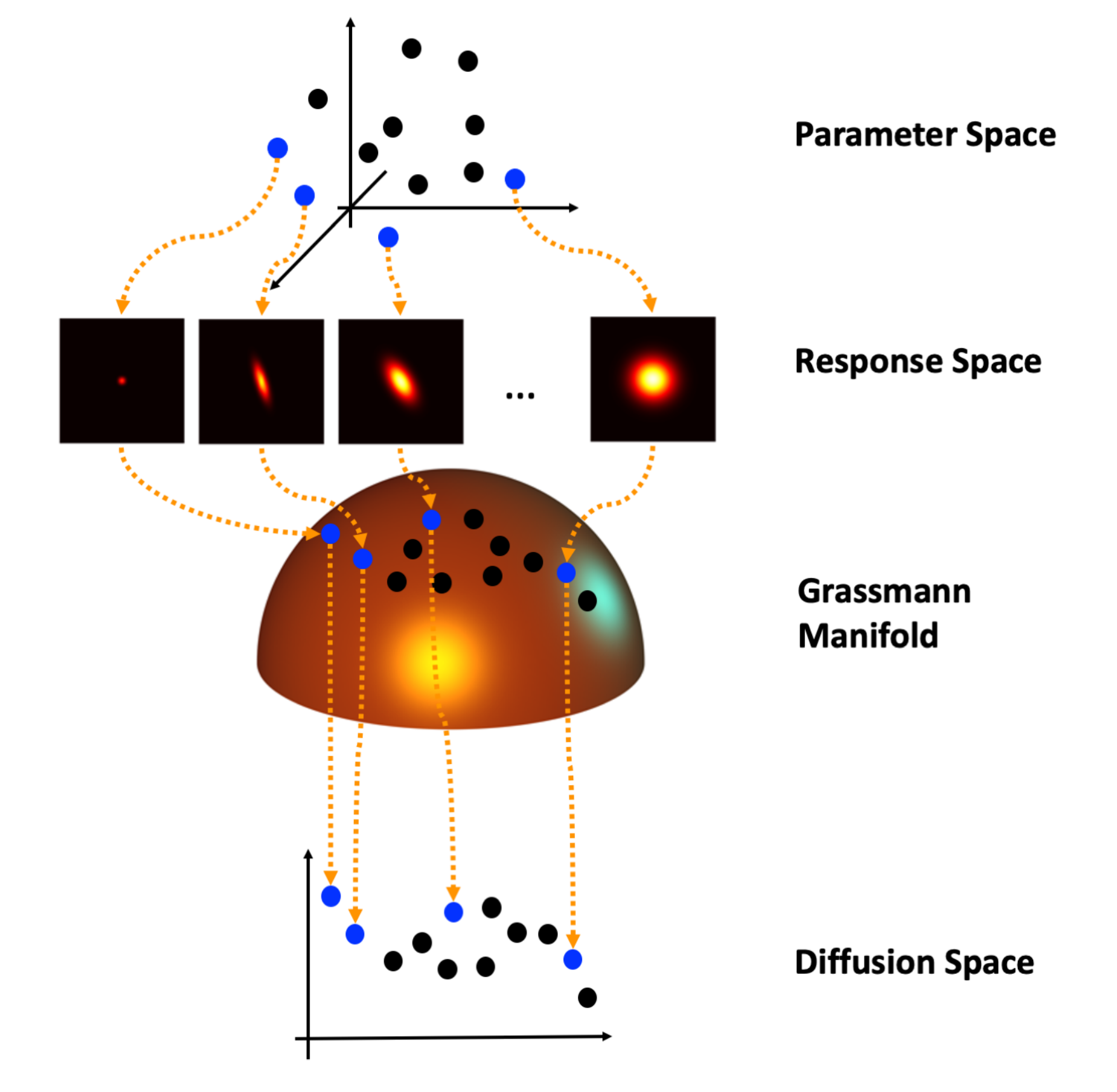}
	\caption{Conceptual illustration of the Grassmannian diffusion maps.}
	\label{fig:fig_gdmaps}
\end{figure}

\noindent
The Grassmannian diffusion maps (GDMaps) \cite{santos2020} is a nonlinear dimension reduction technique that uses diffusion maps to learn the low-dimensional structure of a dataset on the Grassmann manifold. GDMaps is a two-stage dimension reduction. The first dimensional reduction is a pointwise projection of the elements of a dataset onto the Grassmann manifold. Next, a connected graph is created across the data on the Grassmann manifold where a random walk is performed to embed the data into a low-dimensional Euclidean space. This dimension reduction is illustrated in Fig. \ref{fig:fig_gdmaps} and the procedure is detailed herein.

Given a set of high-dimensional data $S_{\mathbf{X}} =  \left\{\mathbf{X}_i, \dots, \mathbf{X}_N \right\}$, with $\mathbf{X}_i \in \mathbb{R}^{n \times m}$ possessing a low-rank structure, one can project $\mathbf{X}_i$ onto a Grassmann manifold via Singular Value Decomposition (SVD) such that $\mathbf{X}_i = \mathbf{U}_i\mathbf{S}_i\mathbf{V}_i^T$, with $\mathcal{U}_i=\mathrm{span}\left(\mathbf{U}_i\right) \in \mathcal{G}(p,n)$, and $\mathcal{V}_i=\mathrm{span}\left(\mathbf{V}_i\right) \in \mathcal{G}(p,m)$. Next, considering a positive semi-definite Grassmannian kernel $k: \mathcal{G}(p,.) \times \mathcal{G}(p,.) \rightarrow \mathbb{R}$ (i.e. the Projection or Binet-Cauchy kernel), we construct the kernel matrices $\mathbf{K}_{\mathcal{U}} = [k_{\mathcal{U},ij}] = \left[k(\mathcal{U}_i,\mathcal{U}_j)\right] \in \mathbb{R}^{N \times N}$ and $\mathbf{K}_{\mathcal{V}} = [k_{\mathcal{V},ij}] = \left[k(\mathcal{V}_i,\mathcal{V}_j)\right] \in \mathbb{R}^{N \times N}$. Next, we compile to composite kernel matrix as either $\mathbf{K}=\mathbf{K}_\mathcal{U}+\mathbf{K}_\mathcal{V}$ or $\mathbf{K}=\mathbf{K}_\mathcal{U}\circ\mathbf{K}_\mathcal{V}$ where $\circ$ denotes the Hadamard product. We use this composite kernel to define a random walk over the data on the manifold, denoted $W=\left(\{S_{\mathcal{U}}, S_{\mathcal{V}}\},F_{\mathbf{\Theta}},\mathbf{P}\right)$, where $S_{\mathcal{U}} = \left\{\mathcal{U}_1, \dots, \mathcal{U}_N \right\}$ and $S_{\mathcal{V}} = \left\{\mathcal{V}_1, \dots, \mathcal{V}_N \right\}$, $F_{\mathbf{\Theta}}$ is the joint probability distribution of the input parameters, and $\mathbf{P}$ is the transition probability matrix. The matrix $\mathbf{P} = [P_{ij}]$ is constructed by first building the following diagonal matrix $\mathbf{D}=[D_{ii}] \in \mathbb{R}^{N \times N}$ as
\begin{equation}\label{eq:5.21}
    D_{ii} = \sum_{j=1}^{N} k_{ij},
\end{equation}
\noindent
such that the stationary distribution of the random walk is given by
\begin{equation}\label{eq:5.22}
    \pi_{i} = \frac{D_{ii}}{\sum_{k=1}^N D_{kk}}.
\end{equation}
\noindent
Next, by normalizing the kernel $k_{ij}$ as follows
\begin{equation}\label{eq:5.23}
    \kappa_{ij} = \frac{k_{ij}}{\sqrt{D_{ii}D_{jj}}}.
\end{equation}
\noindent
the transition matrix $\mathbf{P}$ can be constructed as
\begin{equation}\label{eq:5.24}
    P_{ij} = \frac{\kappa_{ij}}{\sum_{k=1}^{N} \kappa_{ik}}.
\end{equation}
\noindent
The eigendecomposition of $\mathbf{P}$ yields a set of eigenvectors $\mathbf{\Phi} = [\phi_0, \dots, \phi_N]$ and their respective eigenvalues $\mathbf{\Lambda} = \{\lambda_0, \dots, \lambda_N\}$. Thus, every element $\mathbf{X}_i$ of $S_{\mathbf{X}}$ has a representation on a low-dimensional Euclidean space defined by the points $\boldsymbol{\xi}_i = [\lambda_0 \Phi_{ir}, \dots, \lambda_r \Phi_{ir}]^T$, where $r<N$ due to the decaying spectrum $\{\lambda_0, \dots, \lambda_N\}$.

\section{Geometric Harmonics}
\label{s:gh}
The construction of a mapping between the reduced (i.e., diffusion space) and ambient (i.e., parameter space, response space) space has been discussed in the literature by several authors \cite{nadler2006,coifman2008,erban2007,chiavazzo2014,erichson2018}. In this regard, the construction of the lifting (from the reduced space to the ambient space) and restriction (from the ambient space to the reduced space) operators relies on the extension of empirical functions only known at specific locations of the domain, also known as out-of-sample extension.

For example, the Nystr\"{o}m extension, which is very closely related to GP regression \cite{bermanis2013}, is commonly used to construct the restriction operator within the conventional diffusion maps framework \cite{nadler2006,chiavazzo2014}. In particular, given a new sample $\mathbf{X}_k$ in the dataset $S_{\mathbf{X}}$, the diffusion coordinates can be obtained as follows
\begin{equation}\label{eq:nystrom}
    \xi_i(\mathbf{X}_k) = \lambda^{-1}_k\sum_{\mathbf{X} \in S_{\mathbf{X}}} P(\mathbf{X}_k,\mathbf{X})\xi_i(\mathbf{X}),
\end{equation}
where $P(\cdot,\cdot)$ is the transition matrix. 
The Nystr\"{o}m extension has some disadvantages such as the computational complexity of the required diagonalization. Moreover, it can become ill-conditioned \cite{bermanis2013}, so instead we leverage a variation on the Nystr\"{o}m extension, known as geometric harmonics \cite{coifman2006gh}, for constructing both the lifting and restriction operators.

The geometric harmonics aim to extend a function $f$ defined on a set $\Omega$ to a set $\overline{\Omega}$ such that $\Omega \subset \overline{\Omega}$. This extension depends on the selection of an appropriate positive semi-definite kernel $k(\cdot,\cdot)$ (such as the Gaussian kernel in Eq. (\ref{eq:gaussian_kernel})), which defines a unique reproducing kernel Hilbert space $\mathcal{H}$ of functions in $\overline{\Omega}$. Therefore, restricting $k(\cdot,\cdot)$ to $\Omega$ one can define an operator $\mathbf{K}:L^2(\Omega,d\mu) \rightarrow \mathcal{H}$, such that
\begin{equation}\label{eq:conv_gh}
    \mathbf{K}f(\overline{\omega}) = \int_{\Omega}k(\overline{\omega},\omega)f(\omega)d\mu(\omega),
\end{equation}
where $d\mu$ is a measure with $d\mu<+\infty$, $\omega \in \Omega$, and $\overline{\omega} \in \overline{\Omega}$. A lemma presented by Coifman and Lafon \cite{coifman2006gh} shows that the adjoint operator $\mathbf{K}^{*}: \mathcal{H} \rightarrow L^2(\Omega,d\mu)$ is in fact the restriction operator. Moreover, as this operator is self-adjoint and compact, its eigendecomposition exists and one can write the Geometric Harmonics as
\begin{equation}\label{eq:conv_gh}
    \psi_i(\overline{\omega}) = \lambda^{-1}_i\int_{\Omega}k(\overline{\omega},\omega)\psi(\omega)d\mu(\omega),
\end{equation}
which can be summarized by the following expressions
\begin{equation}\label{eq:conv_gh_mat}
    \mathbf{K}\psi_i = \lambda_i \overline{\psi}_i,
\end{equation}
and
\begin{equation}\label{eq:conv_gh_mat2}
    \mathbf{K}^{*}\overline{\psi}_i = \psi_i,
\end{equation}
where $L^2_{\delta} = \mathrm{span}\{\psi_i,i\in D_{\delta}\}$ and $\mathcal{H}_{\delta} = \mathrm{span}\{\overline{\psi}_i,i\in D_{\delta}\}$, with $D_{\delta} = \{i,\lambda_i \geq \delta \lambda_0\}$ and $\delta>0$. Therefore, the mechanization of the extension algorithm is given by two main steps \cite{coifman2006gh}. First, $f$ is projected onto $L^2_{\delta} = \mathrm{span}\{\psi_i,i\in D_{\delta}\}$. Such that,
\begin{equation}\label{eq:conv_gh_ext1}
    f \mapsto \mathbf{P}_{\delta} f = \sum_{j \in D_{\delta}}\langle f, \psi_j \rangle_{\Omega}\psi_j.
\end{equation}
Second, the extension $\mathbf{E}f$ is given by
\begin{equation}\label{eq:conv_gh_ext2}
    \mathbf{E} f(\overline{\omega}) = \sum_{j \in D_{\delta}}\langle f, \psi_j \rangle_{\Omega}\overline{\psi}_j(\overline{\omega}).
\end{equation}
Details on implementation of GH can be found in Algorithm \ref{alg:gh}. In this algorithm, one can observe that the sets $S_x$ and $S_y$ as well as $S_x^{*}$ and $S_y^{*}$ are composed of column vectors. Therefore, if the Euclidean space of interest is the space of matrices, one can transform an element of this space into vectors by stacking its columns. Further, one can easily observe that the restriction operator between a point on the Grassmann manifold $\mathcal{G}(p,n)$ and the Grassmannian diffusion space exists since appropriate kernels can be defined on the Grassmann manifold, see Section \ref{S:GDMAPs}. On the other hand, there is no guarantee that the inverse map (lifting operator) exists due to the orthogonality constraints of the Grassmann manifold. However, assuming that the Grassmann manifold is locally approximated by a flat inner-product space (i.e. the tangent space $\mathcal{T}_{\mathcal{X}_0}\mathcal{G}(p,n)$) constructed at $\mathcal{X}_0$, one can define a local lifting operator from the Grassmannian diffusion space defined on a set $\Omega \in \mathbb{R}^k$ to $\mathcal{T}_{\mathcal{X}_0}\mathcal{G}(p,n)$. It is then straightforward to apply the exponential mapping to project the extended sampling points in $\mathcal{T}_{\mathcal{X}_0}\mathcal{G}(p,n)$ onto $\mathcal{G}(p,n)$ (see Section 2.1).

\begin{algorithm}[h]
\caption{Geometric Harmonics}
\label{alg:gh}
\begin{algorithmic}[1]
\REQUIRE The set $Sx=\{\mathbf{X}_1, \dots, \mathbf{X}_N\} \subset \mathbb{R}^{n}$ represented by the matrix $\mathbf{S}x=\left[\mathbf{X}_1, \dots, \mathbf{X}_N\right]^T \subset \mathbb{R}^{N \times n}$, the target set $Sy=\{\mathbf{Y}_1, \dots, \mathbf{Y}_N\} \subset \mathbb{R}^{m}$ represented by the matrix $\mathbf{S}y=\left[\mathbf{Y}_1, \dots, \mathbf{Y}_N\right]^T \subset \mathbb{R}^{N \times m}$, a positive semi-definite kernel $k(\cdot,\cdot)$, and a new subset $S_{x^{*}}=\{\mathbf{X}^{*}_1, \dots, \mathbf{X}^{*}_N\} \subset \mathbb{R}^{n}$ of $S_x$ represented by the matrix $\mathbf{S}x^{*}=\left[\mathbf{X}^{*}_1, \dots, \mathbf{X}^{*}_M\right]^T \subset \mathbb{R}^{M \times n}$.
\STATE Compute the kernel matrix $\mathbf{K}=[k(\mathbf{X}_i,\mathbf{X}_j)]$.
\STATE Kernel eigendecomposition: $\mathbf{K}\overline{\psi}_i = \overline{\lambda}_i\overline{\psi}_i$, with $i=1,\dots,r$, with $r \leq N$. Eigenvectors and eigenvalues can be written as the matrices $\overline{\mathbf{\Psi}} \in \mathbb{R}^{N \times r}$ and $\overline{\mathbf{\Lambda}} = \mathrm{diag}(\lambda_1, \dots, \lambda_N) \in \mathbb{R}^{r \times r}$, respectively.
\STATE Compute $\mathbf{B} = \overline{\mathbf{\Psi}} \overline{\mathbf{\Lambda}}^{-1}\overline{\mathbf{\Psi}}^{T}\mathbf{S}_y \in \mathbb{R}^{N \times m}$.
\STATE Compute the kernel values for the new element: $\hat{\mathbf{K}}=[k(\mathbf{X}^{*}_i,\mathbf{X}_j)] \in \mathbb{R}^{M \times N}$.
\STATE Extension: $\mathbf{S}_{y^{*}}=\hat{\mathbf{K}}\mathbf{B} \in \mathbb{R}^{M \times m}$.
\ENSURE a new set $\mathbf{S}_{y^{*}}$.
\end{algorithmic}
\end{algorithm}

\section{Grassmannian-Geometric Harmonics Maps}
\label{S:PM}
Consider the random vector $\mathbf{\Theta} \in \mathbb{R}^Q$ having joint probability distribution $F_{\mathbf{\Theta}}\left( \Theta_1, \dots, \Theta_Q\right)$ as the input parameters to a model $\mathcal{M}(\cdot)$. One can obtain samples $\mathbf{\Theta}_i$ as elements of a set $S_{\mathbf{\Theta}} = \left\{\mathbf{\Theta}_1, \dots, \mathbf{\Theta}_N \right\} \subset \Pi$ 
from $F_{\mathbf{\Theta}}\left( \Theta_1, \dots, \Theta_Q\right)$, where $\Pi$ is the parameter space. For each element of $S_{\mathbf{\Theta}}$, the model $\mathcal{M}(\cdot)$ (e.g., finite element model) produces a high-dimensional response $\mathbf{X}_i \in \mathbb{R}^{n \times m}$ such that $\mathbf{X}_i = \mathcal{M}(\mathbf{\Theta}_i)$. Therefore, a set $S_{\mathbf{X}} = \left\{\mathcal{M}(\mathbf{\Theta}_1), \dots, \mathcal{M}(\mathbf{\Theta}_N) \right\} =  \left\{\mathbf{X}_i, \dots, \mathbf{X}_N \right\} \subset \Xi$ is obtained, where $\Xi$ is the response space. With the set $S_{\mathbf{X}}$ and assuming that $\mathbf{X}_i$ has a low-rank structure, we begin by projecting $\mathbf{X}_i$ onto a Grassmann manifold. This operation is performed via singular value decomposition (SVD), as presented in Section \ref{s:grassmann}. Thus, one can decompose $\mathbf{X}_i$ as $\mathbf{X}_i = \mathbf{U}_i\mathbf{S}_i\mathbf{V}_i^T$, with $\mathcal{U}_i=\mathrm{span}\left(\mathbf{U}_i\right) \in \mathcal{G}(p,n)$, and $\mathcal{V}=\mathrm{span}\left(\mathbf{V}\right) \in \mathcal{G}(p,m)$. Moreover, a set $S_{\mathbf{S}} = \left\{\mathbf{S}_1, \dots, \mathbf{S}_N \right\} \subset \Sigma$ of singular values is obtained, where $\Sigma$ is the space of singular values and $\mathbf{S}_i \in \mathbb{R}^{p \times p}$.

Selecting an appropriate Grassmannian kernel \cite{santos2020}, we next construct a connected graph on the sets $S_{\mathcal{U}} = \left\{\mathcal{U}_1, \dots, \mathcal{U}_N \right\} \subset \mathcal{G}(p,n)$ and $S_{\mathcal{V}} = \left\{\mathcal{V}_1, \dots, \mathcal{V}_N \right\} \subset \mathcal{G}(p,m)$ and apply the procedure of Section \ref{S:GDMAPs} to determine the new coordinates (Grassmannian diffusion coordinates) embedding the data on the Grassmann manifolds $\mathcal{G}(p,n)$ and $\mathcal{G}(p,m)$ into a low-dimensional Euclidean space (Grassmannian diffusion space). Once the Grassmannian diffusion coordinates $\boldsymbol{\xi}=\left\{ \boldsymbol{\xi}_1, \dots, \boldsymbol{\xi}_N \right\} \in \Delta$ are obtained, we construct a global map (surrogate) using geometric harmonics (see Section \ref{s:gh}) between $\Pi$ and $\Delta$ ($GH_0: \Pi \rightarrow \Delta$), although local maps can also be constructed if necessary. For this mapping, the Gaussian kernel $k(\mathbf{\Theta}_i,\mathbf{\Theta}_j)$ (Eq. \ref{eq:gaussian_kernel}) is used because we want to construct a map between Euclidean spaces. A second map between $\Pi$ and $\Sigma$ ($GH_1: \Pi \rightarrow \Sigma$) is constructed using GH as well. With both maps constructed, we can predict the dimension reduced response of $\mathcal{M}(\cdot)$ for any new set of input parameters. In other words, considering that a new set of input parameters $\mathbf{\Theta}^{*}$ is sampled from $F_{\mathbf{\Theta}}$, we estimate the corresponding Grassmannian diffusion coordinates in $\Delta$ by $\boldsymbol{\xi}^{*}=GH_0(\mathbf{\Theta}^{*})$. Simultaneously, we estimate the singular values by $\boldsymbol{S}^{*}=GH_1(\mathbf{\Theta}^{*})$. 

Once the estimated coordinate $\boldsymbol{\xi}^{*}$ in $\Delta$ is obtained, it is necessary to expand this reduced dimension solution from the Grassmannian diffusion manifold back to the full-dimensional solution in the ambient space. The first step of this decoding is to define a mapping between $\Delta$ and the Grassmann manifolds $\mathcal{G}(p,n)$ and $\mathcal{G}(p,m)$. To achieve this, we define a series of local GH lifting operators $\Lambda_0: \Delta \rightarrow \mathcal{T}_{\hat{\mu}_u}\mathcal{G}(p,n)$ and $\Lambda_1: \Delta \rightarrow \mathcal{T}_{\hat{\mu}_v}\mathcal{G}(p,m)$ where $\hat{\mu}_u$ and $\hat{\mu}_v$ are the reference points on the Grassmann manifold where the tangent spaces are constructed. The local operators are determined by identifying the $k$ nearest neighbors to the point $\boldsymbol{\xi}^{*}$. These points define the vicinity of $\boldsymbol{\xi}^{*}$ on the Grassmann manifold, and the tangent space can be constructed either around their Karcher mean (see Section \ref{s:gr_km}); or around the nearest neighbor of $\boldsymbol{\xi}^{*}$ on the Grassmann manifold, which is a computationally efficient method since no optimization is performed. These local data are then used to construct the GH lifting operator. Using these lifting operators, we obtain the points $\mathbf{\Gamma}^{*}_{u} = \Lambda_0(\boldsymbol{\xi}^{*})$ and $\mathbf{\Gamma}^{*}_{v} = \Lambda_1(\boldsymbol{\xi}^{*})$, where $\mathbf{\Gamma}^{*}_{u} \in \mathbb{R}^{m \times p}$ and $\mathbf{\Gamma}^{*}_{u} \in \mathbb{R}^{m \times p}$ represent points on their respective tangent spaces. Next, we apply the exponential map to obtain the corresponding points on $\mathcal{G}(p,n)$ and $\mathcal{G}(p,m)$ (see Section \ref{s:exp_log}) as $\mathbf{U}^{*} = \mathrm{exp}_{\hat{\mu}_u}(\mathbf{\Gamma}^{*}_{u})$ and $\mathbf{V}^{*} = \mathrm{exp}_{\hat{\mu}_v}(\mathbf{\Gamma}^{*}_{v})$, where $\mathcal{U}^{*}=\mathrm{span}\left(\mathbf{U}^{*}\right) \in \mathcal{G}(p,n)$ and $\mathcal{V}^{*}=\mathrm{span}\left(\mathbf{V}^{*}\right) \in \mathcal{G}(p,m)$. Finally, the solution $\mathbf{X}^{*}$ for the new set of input parameters $\mathbf{\Theta}^{*}$ can be predicted by the following matrix product

\begin{equation}\label{eq:pred}
    \mathbf{X}^{*} = \mathbf{U}^{*}\mathbf{S}^{*}\mathbf{V}^{*T}.
\end{equation}

\begin{figure}[!ht]
	\centering
	\captionsetup{justification=centering}
	\includegraphics[width=0.6\textwidth]{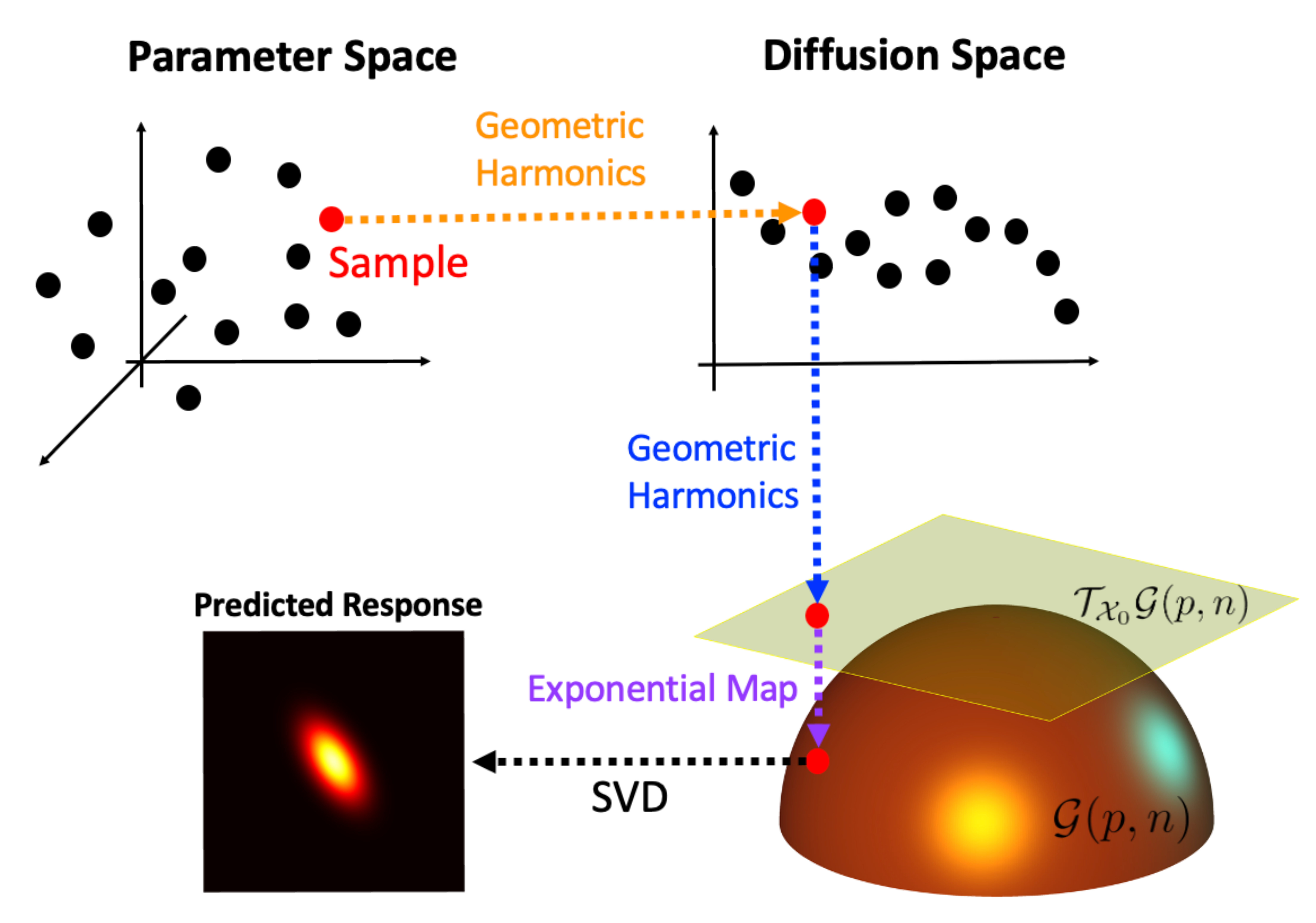}
	\caption{Conceptual illustration of the Grassmannian diffusion maps based surrogate modeling: sampling the parameter space and and mapping to response prediction.}
	\label{fig:figure_prediction}
\end{figure}

Next, two algorithms are presented summarizing this method. Algorithm \ref{alg:gdmaps_train} describes the construction of the maps between the spaces of interest (training), and Algorithm \ref{alg:gdmaps_pred} shows how to predict the response using the constructed maps. Moreover, the proposed surrogate modeling approach is illustrated conceptually in Fig. \ref{fig:figure_prediction}.

\begin{algorithm}
\caption{Grassmannian-GHMaps: training}
\label{alg:gdmaps_train}
\begin{algorithmic}[1]
\REQUIRE a model $\mathcal{M}(\mathbf{\Theta})$; a set of $N$ vectors of input parameters $S_{\mathbf{\Theta}} = \left\{\mathbf{\Theta}_1, \dots, \mathbf{\Theta}_N \right\} \subset \Pi$, with $\mathbf{\Theta} \subset \mathbb{R}^Q$; and a response set $S_{\mathbf{X}} = \left\{\mathcal{M}(\mathbf{\Theta}_1), \dots, \mathcal{M}(\mathbf{\Theta}_N) \right\} =  \left\{\mathbf{X}_i, \dots, \mathbf{X}_N \right\} \subset \Xi$, with $\mathbf{X}_i \in \mathbb{R}^{n \times m}$, of $\mathcal{M}(\mathbf{\Theta}_i)$.
\FOR{$i \in 1, \dots, N$}
\STATE Compute the thin Singular Value Decomposition: $\mathbf{X}_i = \mathbf{U}_i\mathbf{S}_i\mathbf{V}_i^T$, where $\mathcal{U}_i=\mathrm{span}(\mathbf{U}_i) \in \mathcal{G}(p,n)$ and $\mathcal{V}_i=\mathrm{span}(\mathbf{V}_i) \in \mathcal{G}(p,m)$.
\ENDFOR
\STATE Construct the sets $S_{\mathcal{U}} = \left\{\mathcal{U}_1, \dots, \mathcal{U}_N \right\} \subset \mathcal{G}(p,n)$, $S_{\mathcal{V}} = \left\{\mathcal{V}_1, \dots, \mathcal{V}_N \right\} \subset \mathcal{G}(p,m)$, and  $S_{\mathbf{S}} = \left\{\mathbf{S}_1, \dots, \mathbf{S}_N \right\} \subset \Sigma$.
\STATE For every pair $\left[\mathcal{U}_i,\mathcal{U}_j\right]$ and $\left[\mathcal{V}_i,\mathcal{V}_j\right]$ compute the entries of $k_{ij}$ of the kernel matrices $k_{ij}\left(\mathcal{U}\right)$ and $k_{ij}\left(\mathcal{V}\right)$, either using Eq. (\ref{eq:kernel_bc}) or Eq. (\ref{eq:kernel_proj}) (or Eq. (\ref{eq:kernel_bc_angles}) or Eq. (\ref{eq:kernel_proj_angles}), equivalently).
\STATE If necessary, compute the composed kernel matrix $k\left(\mathcal{U},\mathcal{V}\right)$.
\vspace{0.2cm}
$k\left(\mathcal{U},\mathcal{V}\right) = k_{ij}\left(\mathcal{U}\right)+k_{ij}\left(\mathcal{V}\right)$ or $k\left(\mathcal{U},\mathcal{V}\right) = k_{ij}\left(\mathcal{U}\right) \circ k_{ij}\left(\mathcal{V}\right)$, where $\circ$ is the Hadamard product.
\vspace{0.2cm}
\STATE Apply the approach of Section \ref{S:GDMAPs} on $k\left(\mathcal{U},\mathcal{V}\right)$ to get the Grassmannian Diffusion Coordinates $\boldsymbol{\xi}=\left\{ \boldsymbol{\xi}_1, \dots, \boldsymbol{\xi}_N \right\} \in \Delta$.
\STATE Construct the GH map from $\Pi$ to $\Delta$, $GH_0: \Pi \rightarrow \Delta$. (Algorithm 1 with $(\mathbf{\Theta}, \mathbf{\xi})$ as training data.)
\STATE Construct the GH map from $\Pi$ to $\Sigma$, $GH_1: \Pi \rightarrow \Sigma$. (Algorithm 1 with $(\mathbf{\Theta}, \mathbf{S})$ as training data.)
\ENSURE GH maps: $GH_0$ and $GH_1$.
\end{algorithmic}
\end{algorithm}

\begin{algorithm}
\caption{Grassmannian-GHMaps: prediction}
\label{alg:gdmaps_pred}
\begin{algorithmic}[1]
\REQUIRE the GH maps $GH_0$ and $GH_1$ (Algorithm \ref{alg:gdmaps_train}), and a new vector of input parameters $\mathbf{\Theta}^{*}$ to predict $\mathbf{X}^{*} = \mathcal{M}(\mathbf{\Theta}^{*})$.
\STATE Estimate the diffusion coordinates $\boldsymbol{\xi}^{*}$ corresponding to $\mathbf{\Theta}^{*}$: $\boldsymbol{\xi}^{*} = GH_0(\mathbf{\Theta}^{*})$.
\STATE Estimate the singular values $\boldsymbol{S}^{*}$ corresponding to $\mathbf{\Theta}^{*}$: $\boldsymbol{S}^{*} = GH_1(\mathbf{\Theta}^{*})$.
\STATE Find the $k$-neighbors (indices $I_k$) of $\boldsymbol{\xi}^{*}$ in $\boldsymbol{\xi}$ to create the subset $K_{\boldsymbol{\xi}}=\left\{\boldsymbol{\xi}_k| k\in I_k \right\}$.
\STATE From the points with indices in $I_k$, find the reference points (e.g., nearest neighbor, Karcher mean) $\hat{\mu}_u$ and $\hat{\mu}_v$ on $\mathcal{G}(p,n)$ and $\mathcal{G}(p,m)$, respectively.
\FOR{$i \in I_k$}
\STATE Map the corresponding points on $\mathcal{G}(p,n)$ and $\mathcal{G}(p,m)$ to the tangent spaces $\mathcal{T}_{\hat{\mu}_u}$ and $\mathcal{T}_{\hat{\mu}_v}$, respectively: $\mathbf{\Gamma}^{(\mathcal{U})}_i = \mathrm{log}_{\hat{\mu}_k}(\mathcal{U}_i)$ and $\mathbf{\Gamma}^{(\mathcal{V})}_i = \mathrm{log}_{\hat{\mu}_k}(\mathcal{V}_i)$.
\ENDFOR
\STATE Use Algorithm \ref{alg:gh} and the set $K_{\boldsymbol{\xi}}$ to create the local maps $\Lambda_0(\cdot)$ and $\Lambda_1(\cdot)$ between $\Delta$ and $\mathcal{T}_{\hat{\mu}_u}$ and $\mathcal{T}_{\hat{\mu}_v}$ in the vicinity of $\hat{\mu}_u$ and $\hat{\mu}_v$, respectively.
\STATE Compute the matrices corresponding to the points on $\mathcal{T}_{\hat{\mu}_u}$ and $\mathcal{T}_{\hat{\mu}_v}$: $\mathbf{\Gamma}^{*}_{u} = \Lambda_0(\boldsymbol{\xi}^{*})$ and $\mathbf{\Gamma}^{*}_{u} = \Lambda_1(\boldsymbol{\xi}^{*})$.
\STATE Use the exponential map to project the points in the tangent spaces onto their respective Grassmann manifolds: $\mathbf{U}^{*} = \mathrm{exp}_{\hat{\mu}_u}(\mathbf{\Gamma}^{*}_{u})$ and $\mathbf{V}^{*} = \mathrm{exp}_{\hat{\mu}_v}(\mathbf{\Gamma}^{*}_{v})$, where $\mathcal{U}^{*}=\mathrm{span}\left(\mathbf{U}^{*}\right) \in \mathcal{G}(p,n)$ and $\mathcal{V}^{*}=\mathrm{span}\left(\mathbf{V}^{*}\right) \in \mathcal{G}(p,m)$.
\ENSURE predicted solution $\mathbf{X}^{*} = \mathbf{U}^{*}\mathbf{S}^{*}\mathbf{V}^{*T}$.
\end{algorithmic}
\end{algorithm}

\section{Examples}
\label{s:examples}
In this section, three examples are considered to demonstrate the versatility of the proposed surrogate modeling approach. We begin with a toy example in which structured points on the Grassmann manifold can be easily visualized as points on the unit sphere. In the second example, the electrical potential field of an infinitely long dielectric cylinder suspended in a homogeneous electric field is predicted considering that the cylinder's radius $r_0$ and the strength of the electric field $E_{\infty}$ are random variables. The third example considers the evolution of the strain field in an amorphous solid under simple shear using the shear transformation zone (STZ) theory of plasticity. 

The projection kernel in Eq. (\ref{eq:kernel_proj}) is adopted in all examples presented in this section, and the kernel composition by the Hadamard product is considered. Moreover, the accuracy of the predicted solutions is evaluated by using the entry-wise relative error for matrices (Eq. (\ref{eq:abs_rel})) and the relative error in a $L_2$-norm (Frobenius for matrices) sense (Eq. (\ref{eq:l2_error})).

\begin{equation}\label{eq:abs_rel}
    \mathrm{error}_{rel} = \Bigg|\frac{\mathbf{X}^{*} - \mathbf{X}_{exact}}{\mathbf{X}_{exact}}\Bigg|.
\end{equation}

\begin{equation}\label{eq:l2_error}
    \mathrm{error}_{L_2} = \frac{\|\mathbf{X}^{*} - \mathbf{X}_{exact}\|_{L_2}}{\|\mathbf{X}_{exact}\|_{L_2}}.
\end{equation}

\subsection{Structured data on the unit sphere in $\mathbb{R}^3$}
\label{ex:1}
\begin{figure}[!ht]
	\centering
	\captionsetup{justification=centering}
	\includegraphics[width=\textwidth]{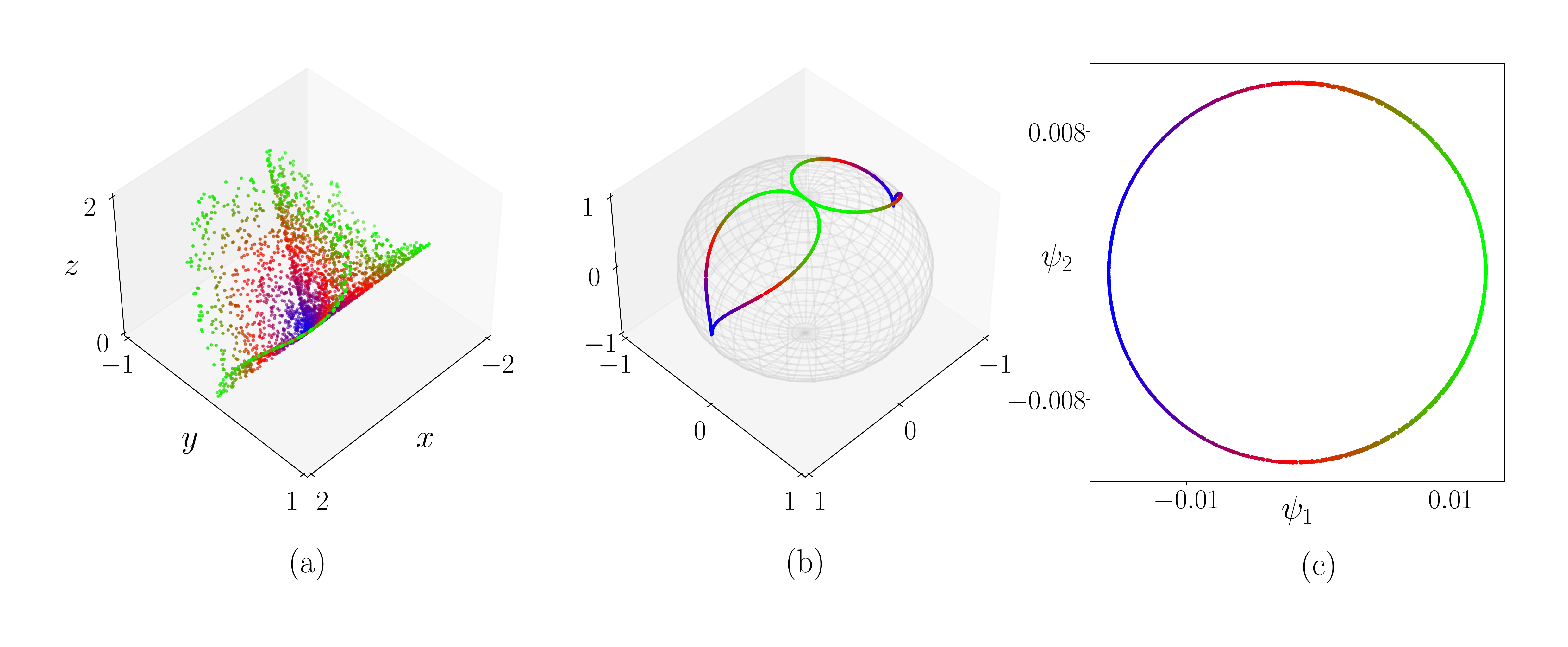}
	\caption{Example 1: Collection of $N=3,000$ random points constrained by Eqs. (29) (a) in the ambient space $\mathbb{R}^3$, (b) in $\mathcal{G}(1,3)$ or $\mathbb{S}^3$, (c) in Grassmannian diffusion coordinates. In (a) points are colored by Euclidean distance from the origin. In (b), (c) points are colored by the first Grassmannian diffusion coordinate.}
	\label{fig:ex1_cones}
\end{figure}

Consider the following set of equations,
\begin{equation} \label{eq1}
\begin{split}
x & = |r| \mathrm{sin}(t)\mathrm{cos}(s), \\
y & = r \mathrm{sin}(t)\mathrm{sin}(s), \\
z & = |r| \mathrm{cos}(t);
\end{split}
\end{equation}
such that $r$ is uniformly distributed in the interval $[-2,2]$, $t$ is uniformly distributed in the interval $[-\pi/2,\pi/2]$, and $s=\mathrm{sin}^{-1}\left(\mathrm{cos}(t)^2\right)$. We draw $N=3,000$ sample pairs $(r,t) \in \Pi$ to obtain a collection of $N$ points constrained on two cone-like structures in $\mathbb{R}^3$ as presented in Fig. \ref{fig:ex1_cones}a, with the colors representing the magnitude $\sqrt{x^2 + y^2 + z^2}$. In effect, we have a model (Eq. \eqref{eq1}) that maps two random variables onto a surface in $\mathbb{R}^3$.

Each point is represented by a column vector $\mathbf{X}_i = [x_i, y_i, z_i]^T$, which together compose the set $S_{\mathbf{X}} = \left\{\mathbf{X}_1, \dots, \mathbf{X}_N \right\}$. One can readily see that these points can be projected onto the Grassmann manifold $\mathcal{G}(1,3)$, which is the unit sphere $\mathbb{S}^3$. A point $\mathcal{X}_i$ on $\mathcal{G}(1,3)$ is given by the unit vector obtained from the normalization of $\mathbf{X}_i$ such that $\mathcal{X}_i = \mathbf{X}_i/\|\mathbf{X}_i\|_2$, which reveals two inverted teardrop shaped structures on the sphere as illustrated in Fig. \ref{fig:ex1_cones}b. Applying diffusion maps to these points on the Grassmann manifold, we see that a well-defined parametrization is obtained, as revealed by the Grassmannian diffusion coordinates in Fig. \ref{fig:ex1_cones}c.

\begin{figure}[!ht]
	\centering
	\captionsetup{justification=centering}
	\includegraphics[scale=0.4]{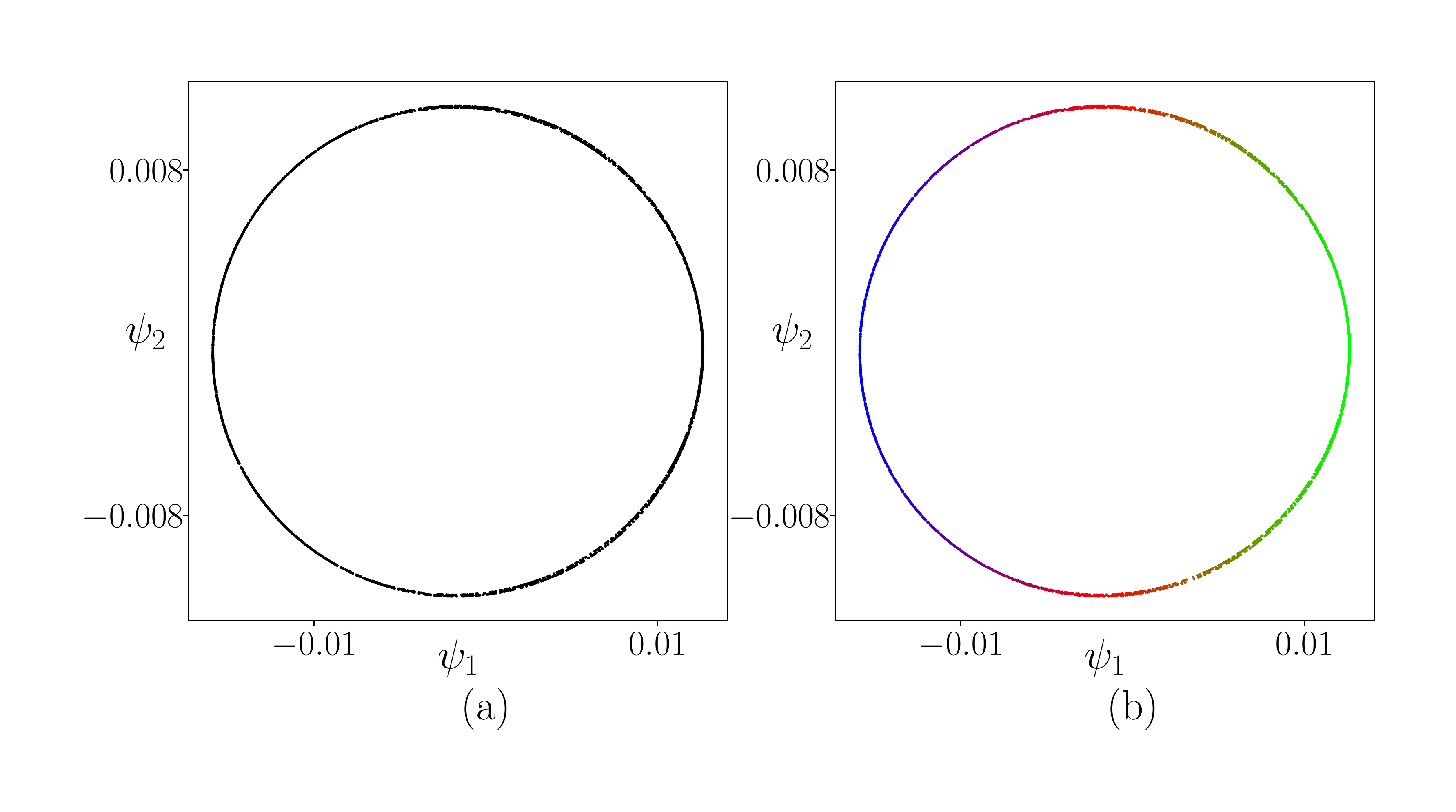}  
	\caption{Example 1: Grassmannian diffusion manifold: a) training set for GH, and b) predicted Grassmannian diffusion manifold for 3,000 additional samples.}
	\label{fig:ex1_samples_gh}
\end{figure}

Geometric harmonics is used to create a map $GH_0:\Pi \rightarrow \Delta$ from the parameter space $\Pi$ to the Grassmannian diffusion manifold $\Delta$. Therefore, it can be considered as a manifold learning technique, where the position on the Grassmannian diffusion manifold (Fig. \ref{fig:ex1_cones}c) can be predicted for any point $\mathbf{\Theta}$ in the parameter space $\Pi$. To verify the accuracy of this learning process, we draw 3,000 additional samples $\mathbf{\Theta}$ (Fig. \ref{fig:ex1_samples_gh}). One can easily see in Fig. \ref{fig:ex1_samples_gh} that the trained GH can reliably predict the shape of the Grassmannian diffusion manifold.

\begin{figure}[!ht]
	\centering
	\captionsetup{justification=centering}
	\includegraphics[width=\textwidth]{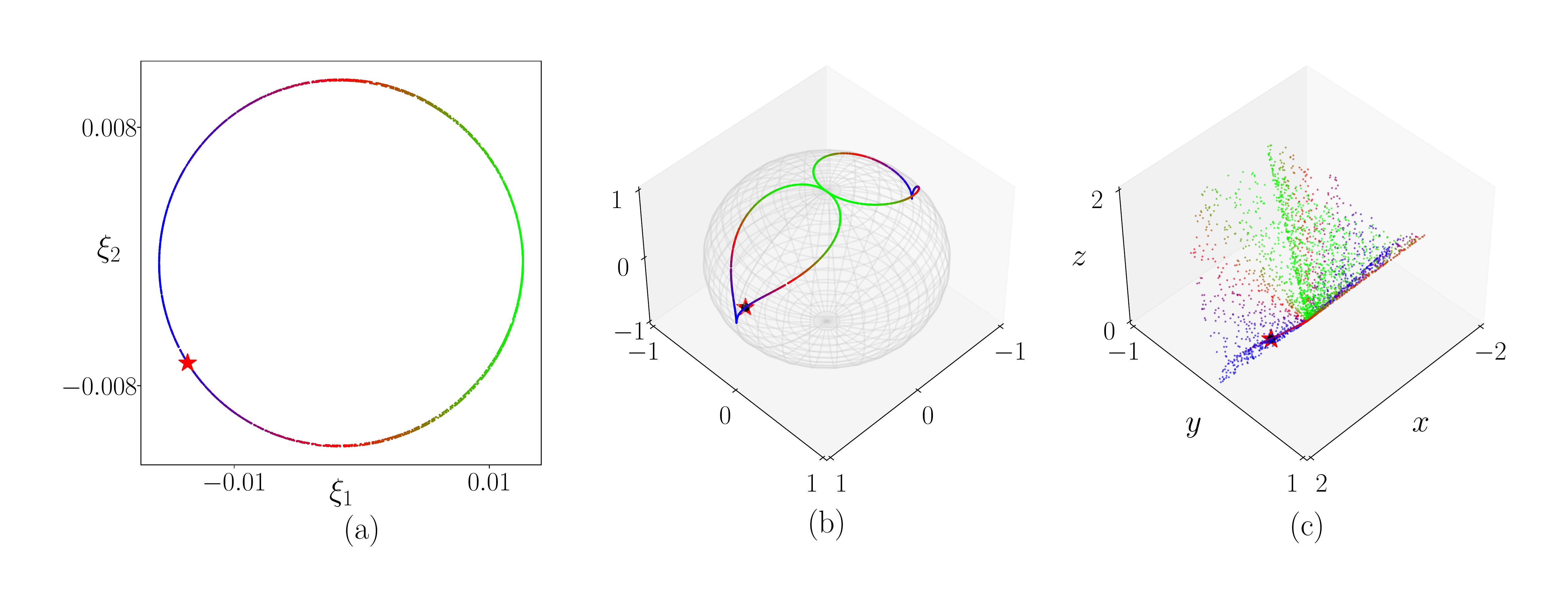}
	\caption{Example 1: Grassmannian diffusion coordinates with the predicted point in the Grassmannian diffusion space.}
	\label{fig:ex1_newgdmaps}
\end{figure}

We obtain a new parameter vector $\mathbf{\Theta}^{*}=(r,t)=(1,1.3)$ by sampling $\Pi$, and the GDMaps-based surrogate model is used to predict the vector $\mathbf{X}^{*}$. Using the map $GH_1:\Pi \rightarrow \Sigma$ one can predict the first two nontrivial diffusion coordinates as $\boldsymbol{\xi}^{*}=[-1.3657 \times 10^{-2},  -6.5931 \times 10^{-3}]$, represented by the red star in Fig. \ref{fig:ex1_newgdmaps}a. Moreover, we observe that $\mathrm{\boldsymbol{S}}^{*} = r = 1$ determines the magnitude of the predicted point in $\mathbb{R}^3$. 
Selecting the $k=3$ nearest neighbors of $\boldsymbol{\xi}^{*}$ in Algorithm \ref{alg:gdmaps_pred}, we predict the point in the ambient space (i.e. on the cone structure) corresponding to $\mathbf{\Theta}^{*}$ by mapping $\boldsymbol{\xi}^{*}$ onto the tangent space $\mathcal{T}$, constructed in the closest neighbor of $\boldsymbol{\xi}^{*}$ (one can also use the Karcher mean alternatively), for posterior projection onto the Grassmann manifold $\mathcal{G}(1,3)$, as illustrated by the red star in Figure \ref{fig:ex1_newgdmaps}b. This point coincides very closely with with true point denoted by the black dot. From the the point projected onto the Grassmann manifold and considering the magnitude of the magnitude of the point given by $|r|$, we predict the point $\mathbf{X}^{*} = [0.96117, 0.06895, 0.2675]^T$ represented by the red star in the ambient space in Fig. \ref{fig:ex1_newgdmaps}c, where again the black dot is the true value $\mathbf{X}_{exact} = [0.9611, 0.0689, 0.2675]$. In this case, we obtain $\mathrm{error}_{rel} = [8.5947 \times 10^{-5}, 8.3515 \times 10^{-3}, 5.6326 \times 10^{-4}]$ and $\mathrm{error}_{L_2} = 6 \times 10^{-4}$.

\begin{figure}[!ht]
	\centering
	\captionsetup{justification=centering}
	\includegraphics[width=0.5\textwidth]{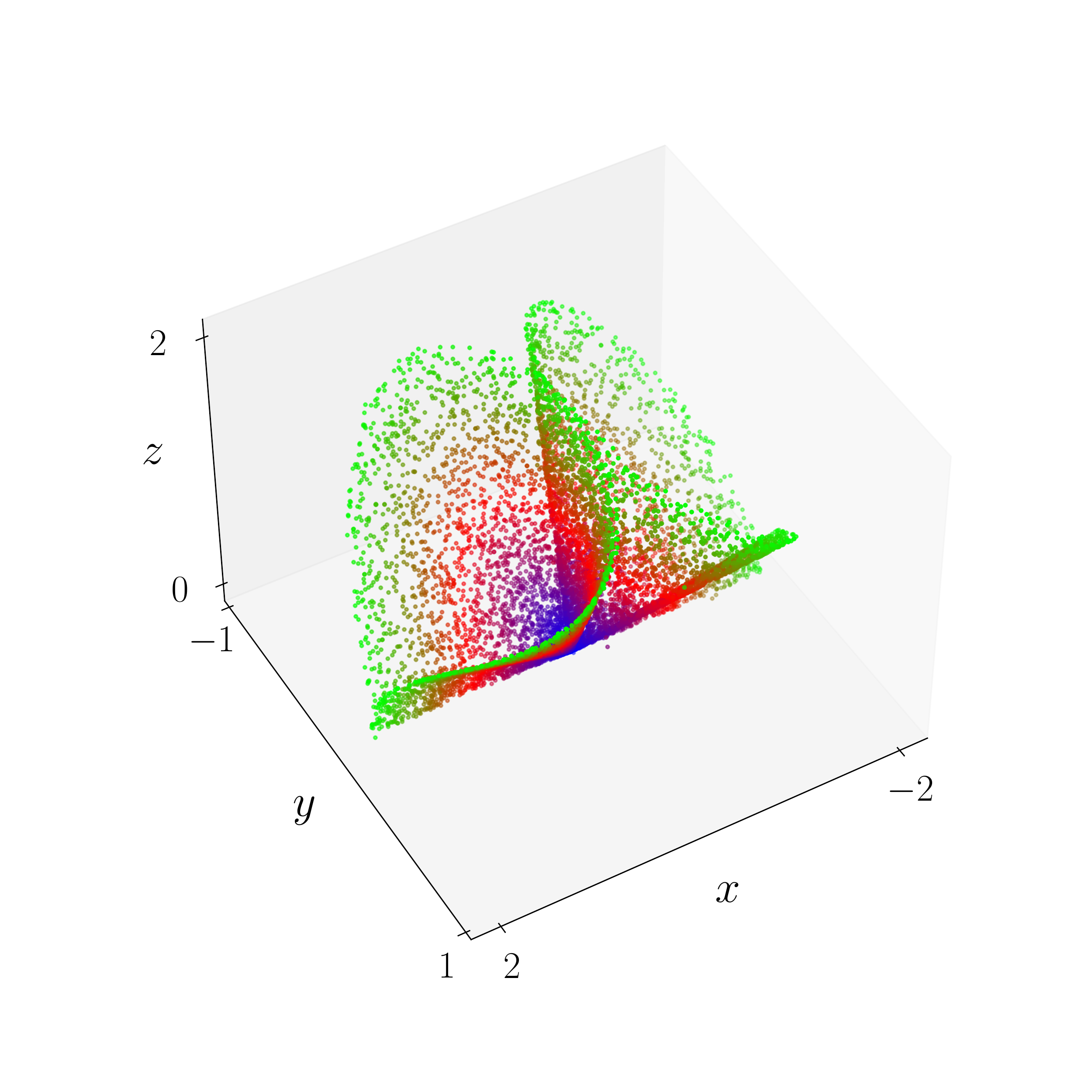}
	\caption{Example 1: Predicted points on the cone-like structure in the ambient space using the surrogate model. Color scale indicates Euclidean distance from the origin.}
	\label{fig:ex1_cones_pred}
\end{figure}

Next, we draw $N=10,000$ pairs $(r,t) \in \Pi$ to assess the overall performance of the proposed surrogate modeling technique. Using this new set of input parameters we predict $N$ points on the cone-like structure and compare them with the exact points corresponding to the set of input parameters. Figure \ref{fig:ex1_cones_pred} shows the predicted cone-like structures from these 10,000 surrogate model evaluations. Comparing with Figure \ref{fig:ex1_cones}, we can see that the points closely match the true structure.

\begin{figure}[!ht]
	\centering
	\captionsetup{justification=centering}
	\includegraphics[width=\textwidth]{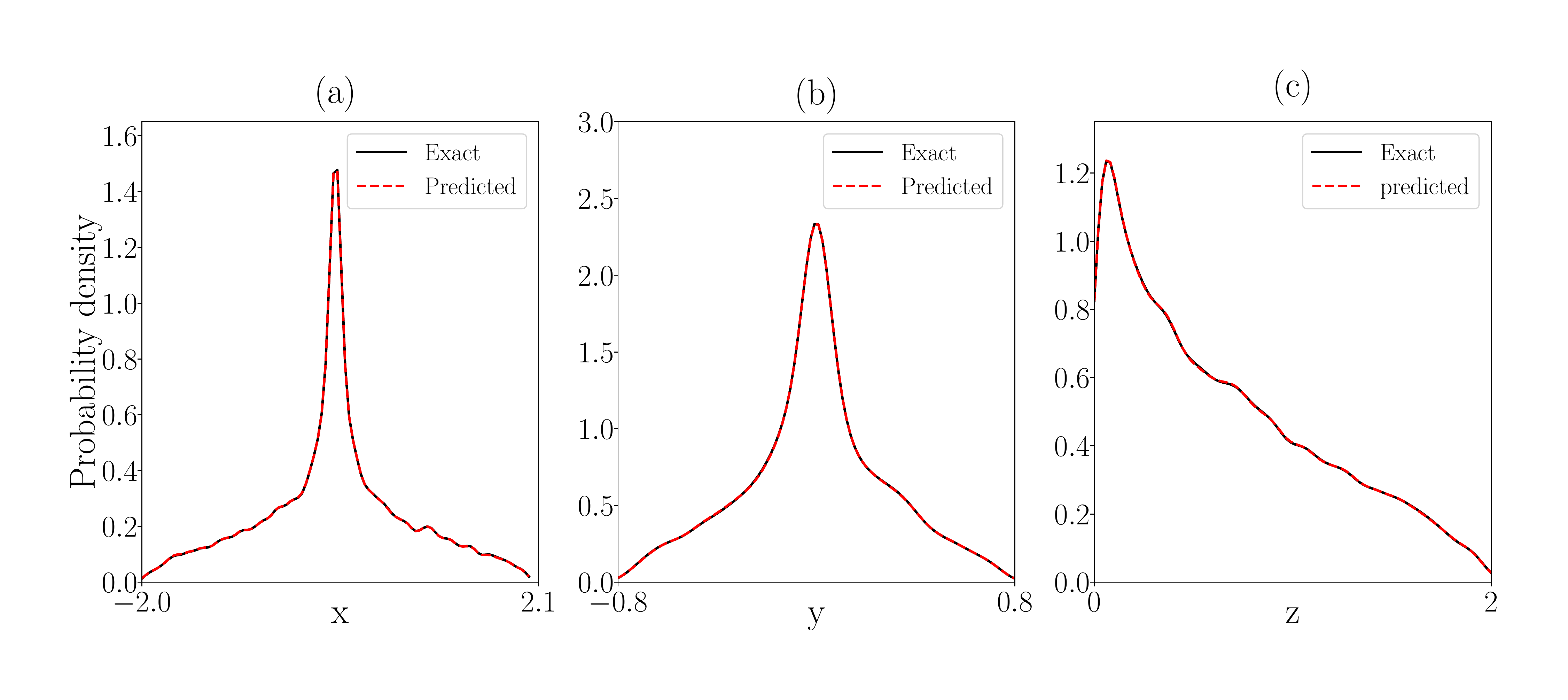}
	\caption{Example 1: Probability density functions for dimensions (a) $x$, (b) $y$, and (c) $z$.}
	\label{fig:ex1_pdfs}
\end{figure}

To assess the overall quality of the predictions, the marginal probability density functions (PDF) for each dimension $(x,y,z)$ are estimated using the kernel density estimation (KDE) and shown in Fig. \ref{fig:ex1_pdfs} for both the true samples and the surrogate predictions. The PDFs for the surrogate predictions match those of the true samples very closely. 

\subsection{Dielectric cylinder in homogeneous electric field}
\label{ex:2}
In this example, we study variations in the electrical potential of an infinitely long dielectric cylinder suspended in a homogeneous electric field, resulting from uncertainty in the input parameters. The problem is defined over a rectangular domain $\Omega = \left[-1,1\right] \times \left[-1,1\right]$ with the embedded cylinder domain $D_{\text{c}} = \left\{\mathbf{x} =\left(x,y\right) \:|\: \sqrt{x^2 + y^2} \leq r_0\right\}$, where $r_0$ is the cylinder's radius. We assume Dirichlet boundary conditions, $\Gamma_\text{D}$, on the left and right boundaries and Neumann boundary conditions $\Gamma_\text{N}$ on the top and bottom boundaries. 

The electric potential $u(\mathbf{x})$ in $\Omega$ can be computed by solving the Laplace equation
\begin{subequations}
	\label{eq:poisson_eq_gen}
	\begin{align}
	-\nabla \cdot \left(\varepsilon \left(\mathbf{x}\right)  \nabla u\left(\mathbf{x}\right)\right)&=0, &&\mathbf{x}\in \Omega, \\
	u\left(\mathbf{x}\right)&=u^*\left(\mathbf{x}\right), &&\mathbf{x} \in \Gamma_{\text{D}}, \\
	\left(\nabla u\left(\mathbf{x}\right)\right)\cdot \mathbf{n}&=\left(\nabla u^*\left(\mathbf{x}\right)\right)\cdot \mathbf{n}, &&\mathbf{x}\in \Gamma_{\text{N}},
	\end{align}
\end{subequations}
where $\mathbf{n}$ denotes the outer normal unit vector. The permittivity $\varepsilon\left(\mathbf{x}\right)$ is given by
\begin{align}
\varepsilon(\mathbf{x})=
\begin{cases}
\varepsilon_{\text{c}}, &\mathbf{x} \in D_{\text{c}}, \\
\varepsilon_{\text{o}}, &\mathbf{x} \in \Omega \setminus D_{\text{c}},
\end{cases}
\end{align} 
and $u^*$, which is also the analytical solution to the problem, is given by
\begin{align}
\label{eq:material_prob}
u^{*}\left(\mathbf{x}\right)=-E_{\infty}x\:
\begin{cases}
1 - \frac{\varepsilon_{\text{c}} / \varepsilon_{\text{o}} - 1}{\varepsilon_{\text{c}} / \varepsilon_{\text{o}} + 1}\frac{r_0^2}{x^2 + y^2}, &\mathbf{x} \in \Omega \setminus D_{\text{c}}, \\
\frac{2}{\varepsilon_{\text{c}} / \varepsilon_{\text{o}} + 1}, &\mathbf{x} \in D_{\text{c}}.
\end{cases}
\end{align}
where $E_\infty$ is the strength of the homogeneous electric field. We consider as random variables the cylinder's radius $r_0$, and the strength of the electric field $E_{\infty}$. These variables are uniformly distributed as defined in Table \ref{table:cylinder}. 

\begin{center}
\begin{table}[!ht]
\centering
\caption{Details of the state variables and input parameters of the dielectric cylinder suspended in homogeneous electric field.}
\vspace{-7pt}
\begin{tabular}{l l c c}
\toprule
Description of variables/parameters & \hspace{15pt} &  & \hspace{2pt} Uncertainty/value \\ [0.5ex]
\toprule
Cylinder radius  &   & $r_0$ & $\sim \mathcal{U}(0.20, 0.70)$\vspace{2pt}  \\ 
Strength of electric field   &   &   $E_{\infty}$ & $\sim \mathcal{U}(8, 18)$ \vspace{2pt} \\
Relative permittivity of cylinder's material   &   &  $\varepsilon_\text{c}$ &    $3$ \vspace{2pt}  \\  
Relative permittivity of surrounding space   &   &  $\varepsilon_\text{o}$ &    $1$  \vspace{1pt} \\ 
\bottomrule
{*All sizes are expressed in SI units}.
\end{tabular}
\label{table:cylinder}
\end{table}
\end{center}


The GDMaps based surrogate model presented in this paper is composed of three distinct maps based on geometric harmonics:  (1) A global map between the parameter space and the Grassmannian diffusion manifold ($GH_0: \Pi \rightarrow \Delta$); (2) a global map between the parameter space and the singular values space ($GH_1: \Pi \rightarrow \Sigma$); and (3) a local map between the Grassmannian diffusion manifold and the tangent space of a region of the Grassmann manifold $\Lambda: \Delta \rightarrow \mathcal{T}$. Moreover, an additional map is built for the projection of the points in the tangent space onto the Grassmann manifold. Considering this sequence of mappings, one can expect that errors can propagate within the framework pipeline, negatively affecting the response prediction. Therefore, the identification of the source of errors is relevant for increasing the accuracy of the predicted outcomes. Therefore, the model presented in this example is used to demonstrate how the errors can be reduced by an appropriate selection of parameters for the construction of maps based on GH.

For the error analysis presented in this section, let's assume a training set of input parameters $\mathbf{\Theta} = (r_{0}, E_{\infty})$ of $N=300$ samples and the corresponding electrical potential fields $\mathbf{X} = \mathcal{M}(\mathbf{\Theta}) \in \mathbb{R}^{m \times n}$ over the domain $\Omega$ discretized into $(n \times m)=(300\times300)=90,000$ mesh points. In this example, the Grassmann manifolds $\mathcal{G}(p,n)=\mathcal{G}(74, 300)$ and $\mathcal{G}(p,m)=\mathcal{G}(74, 300)$ are sufficient to encode the geometric structure of the electrical potential fields $\mathbf{X}$. Using Algorithm \ref{alg:gdmaps_train}, we obtain points on the Grassmannian diffusion manifold; and according to the parsimonious representation of the diffusion maps \cite{dsilva2015}, only the first non-trivial diffusion coordinate is relevant to encode the geometrical information of the underlying physical phenomenon. Therefore, a dimension reduction from $n\times m = 90,000$ to $r=1$ is achieved using Grassmannian diffusion maps. 

\begin{figure}[!ht]
	\centering
	\captionsetup{justification=centering}
	\includegraphics[width=0.65\textwidth]{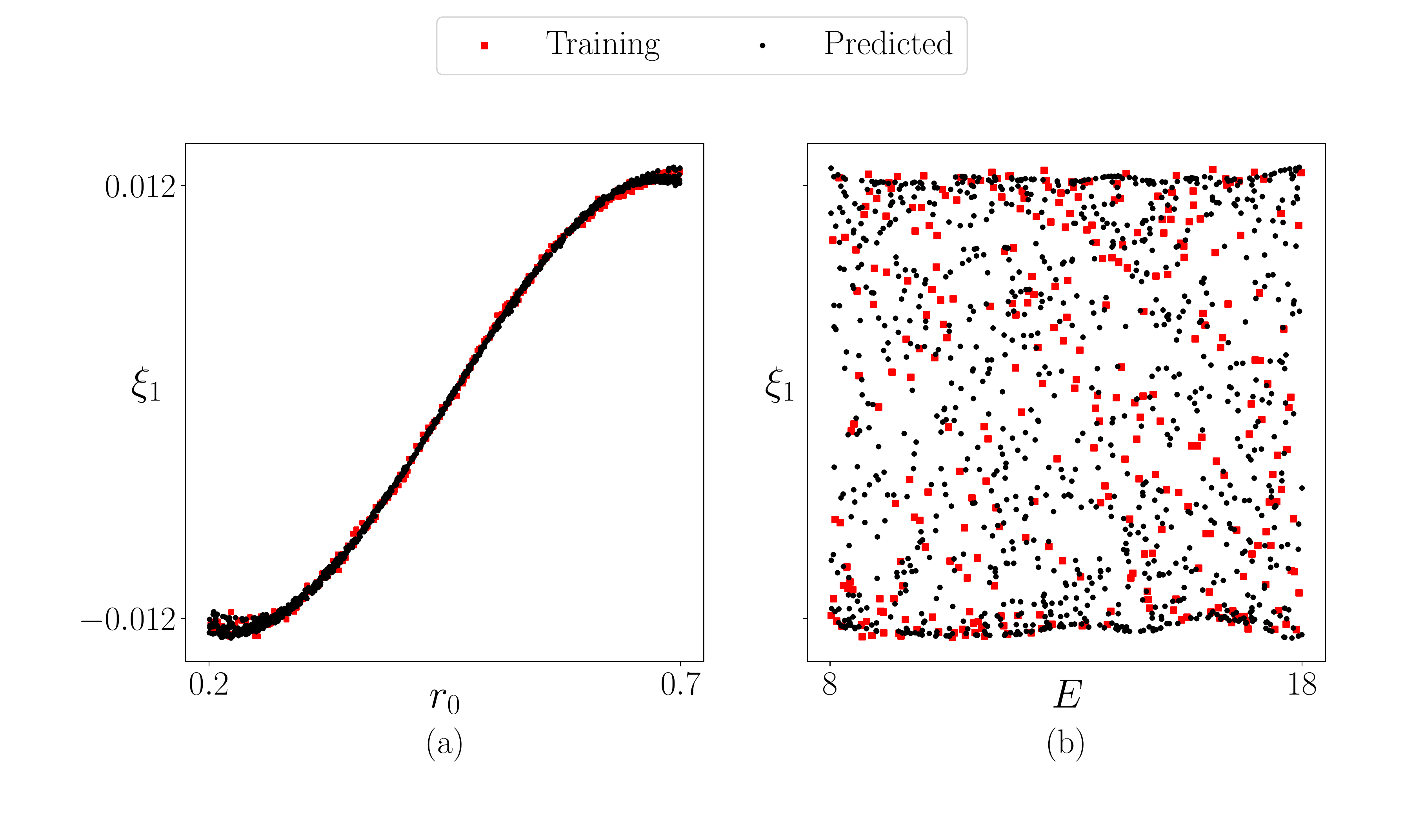}
	\caption{Example 2: the first Grassmannian diffusion coordinate $\xi_1$, and the predicted coordinates of 1,000 additional samples, as a function of a) $r_0$ and b) $E$.}
	\label{fig:ex2_samples_gh}
\end{figure}

To assess the ability of the global map $GH_0: \Pi \rightarrow \Delta$ to learn the relationship between points in the parameter space and the structure of the data on the Grassmannian diffusion manifold ($\Delta$), we start by sampling 1,000 additional points from  $\Pi$ to be mapped in $\Delta$ using $GH_0$. In Fig. \ref{fig:ex2_samples_gh}, which shows both training data and predicted points in the Grassmannian diffusion coordinates, we see that $GH_0$ has adequately learned the shape of the data on the Grassmannian diffusion manifold using Gaussian kernel with length-scale equal to $\epsilon_G=1$, and retaining $q=50$ eigenvalues in the GH framework. However, these parameter, together with the number of samples in the training set, can influence the accuracy of the geometric harmonics maps. Next, the influence of ($\epsilon_G$) and $q$ on the accuracy of both $GH_0: \Pi \rightarrow \Delta$ and $GH_1: \Pi \rightarrow \Sigma$ is analyzed. 

\begin{figure}[!ht]
	\centering
	\captionsetup{justification=centering}
	\includegraphics[width=0.65\textwidth]{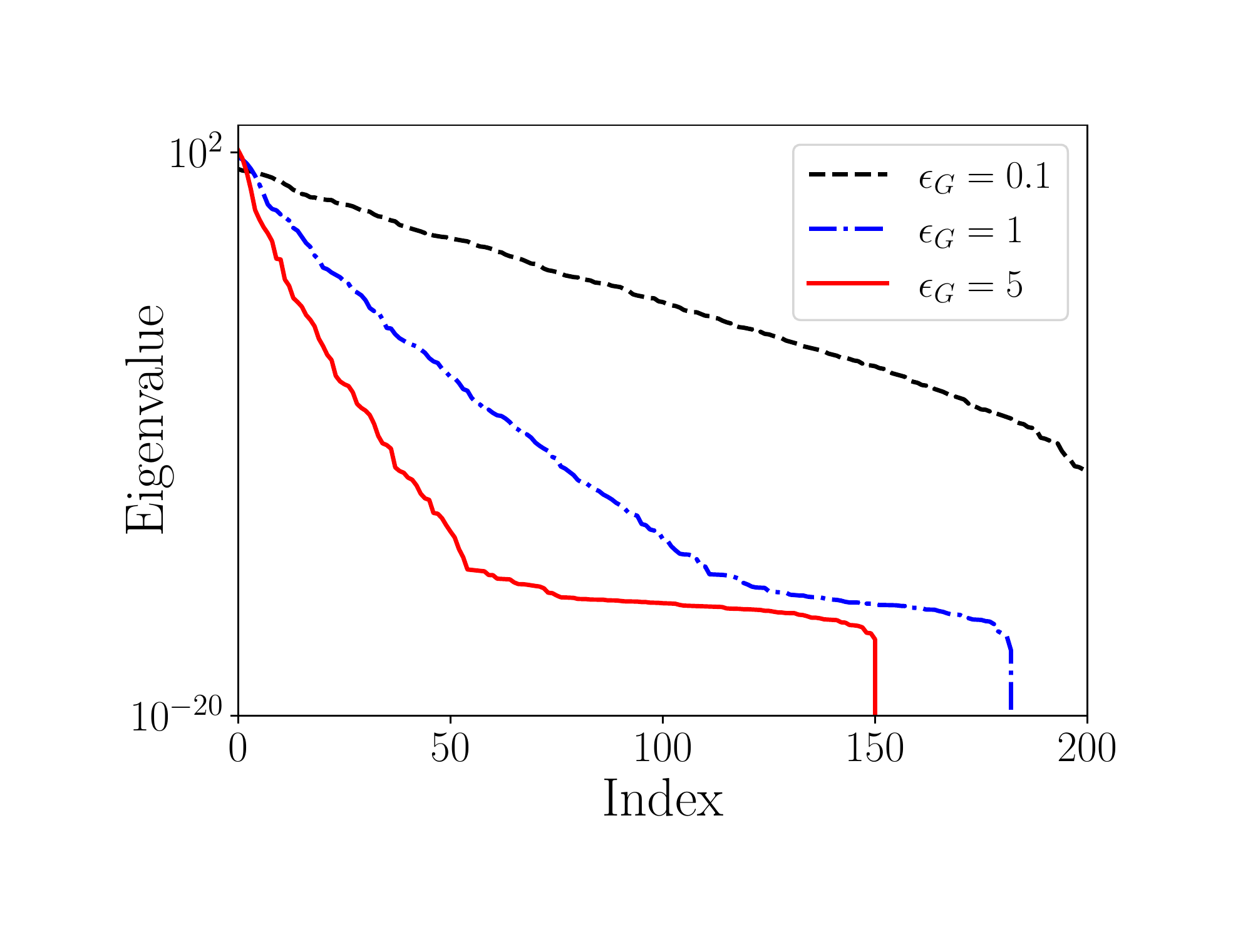}
	\caption{Example 2: Decay of the eigenvalues for the geometric harmonics surrogate $GH_0$ considering different length-scale parameters $\epsilon_G$ in the Gaussian kernel.}
	\label{fig:ex2_eigenvalues}
\end{figure}

\begin{figure}[!ht]
	\centering
	\captionsetup{justification=centering}
	\includegraphics[width=1.\textwidth]{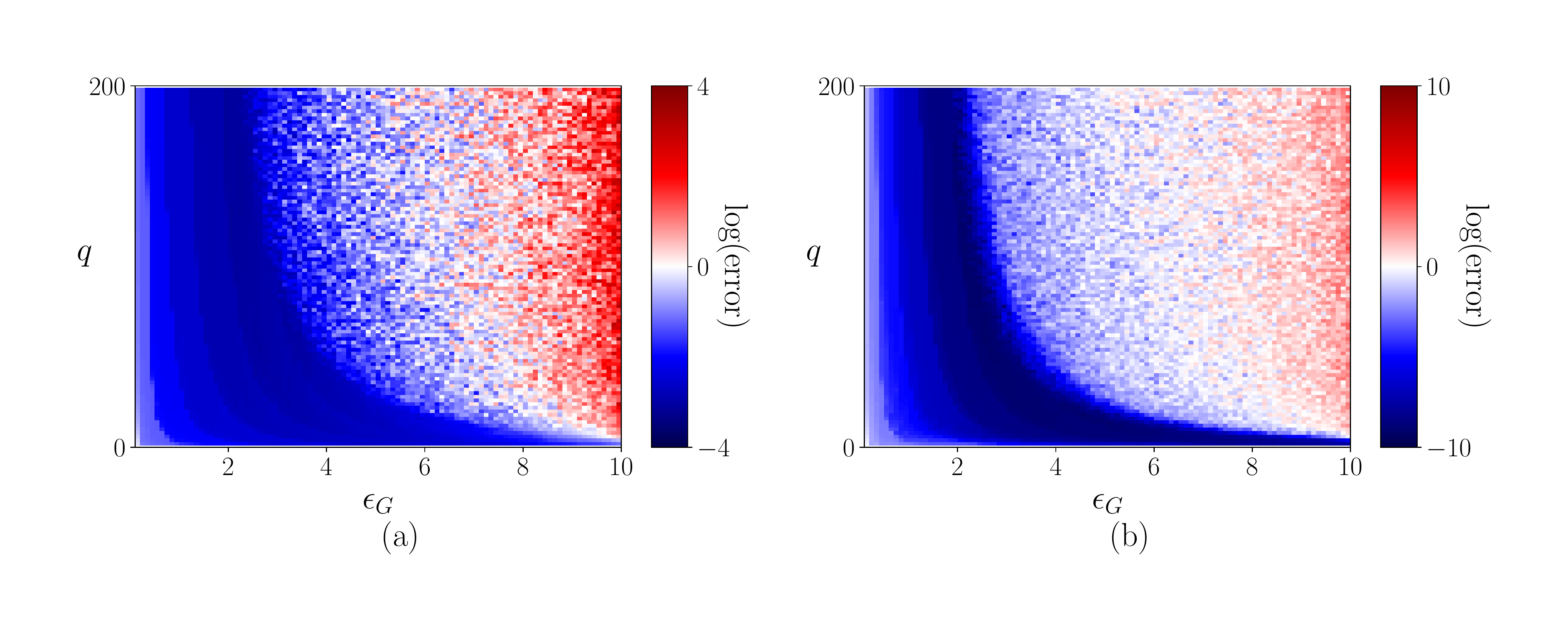}
	\caption{Example 2: Average error from 100 test samples for different combinations of the number of eigenvalues retained in the construction of the geometric harmonics ($q$) and the value to the length-scale parameter ($\epsilon_G$) for (a) surrogate $GH_0$, and (b) surrogate ($GH_1$).}
	\label{fig:ex2_gh0_error}
\end{figure}

We start by plotting the decay of the eigenvalues for $GH_0$ for different values of $\epsilon_G$. It is clear from Fig. \ref{fig:ex2_eigenvalues} that as $\epsilon_G$ increases, the eigenvalues tends to decay quicker. This behavior will have a strong influence on the prediction error of $GH_0$ and $GH_1$ because the construction of the matrix $\boldsymbol{B}$ in Algorithm \ref{alg:gh} depends on the reciprocal of the corresponding eigenvalues. Therefore, if a larger number of very small eigenvalues are retained, their reciprocal could lead to large errors and numerical instabilities. Thus, the selection of $\epsilon_G$ is inherently connected with the number of eigenvalues one should retain for the construction of $\boldsymbol{B}$. This analysis is presented in Fig. \ref{fig:ex2_gh0_error}a for $GH_0$, and in Fig. \ref{fig:ex2_gh0_error}b for $GH_1$ where we show the average error in the GH predictions for different combinations of $\epsilon_G$ and $q$. In both cases, we see that for large values of $\epsilon_G$, a smaller number of eigenvectors and their respective eigenvalues should be retained in the construction of the matrix $\boldsymbol{B}$, which has a direct influence in its rank. Selecting a larger $\epsilon_G$ along with a high $q$ will introduce large errors.

\begin{figure}[!ht]
	\centering
	\captionsetup{justification=centering}
	\includegraphics[width=1.\textwidth]{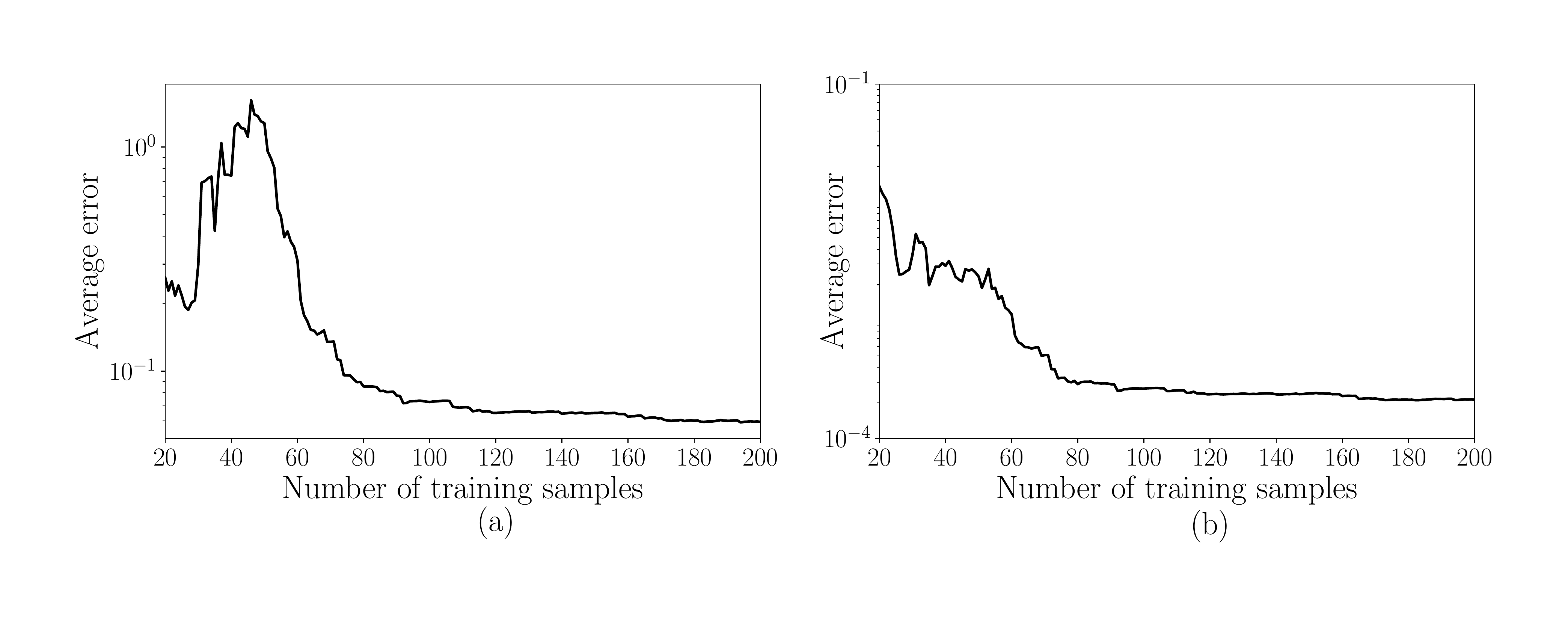}
	\caption{Example 2: Average error from surrogates $GH_0$ (a) and $GH_1$ (b) for increasing size of the training data set..}
	\label{fig:ex2_gh0_error_samp}
\end{figure}

Next, the influence of the size of the training set $N$ is investigated. In this regard, the initial 300 samples and their respective diffusion coordinates are split into a training and a testing set to which the predicted Grassmannian diffusion coordinates can be compared. Considering $N_{train}$ varying from 20 to 200, and keeping $\epsilon_G=1$ and $q=50$ constants, the average error in the $L_2$-norm sense as a function of $N_{train}$ is presented in Fig. \ref{fig:ex2_gh0_error_samp}a, for $GH_0$, and in Fig. \ref{fig:ex2_gh0_error_samp}b, for $GH_1$. In both cases the error reduces, although at a limited rate, after a certain value of $N$ due to the fact that a residual error remains due to the selected values for $\epsilon_G$ and $q$.

\begin{figure}[!ht]
	\centering
	\captionsetup{justification=centering}
	\includegraphics[width=1.\textwidth]{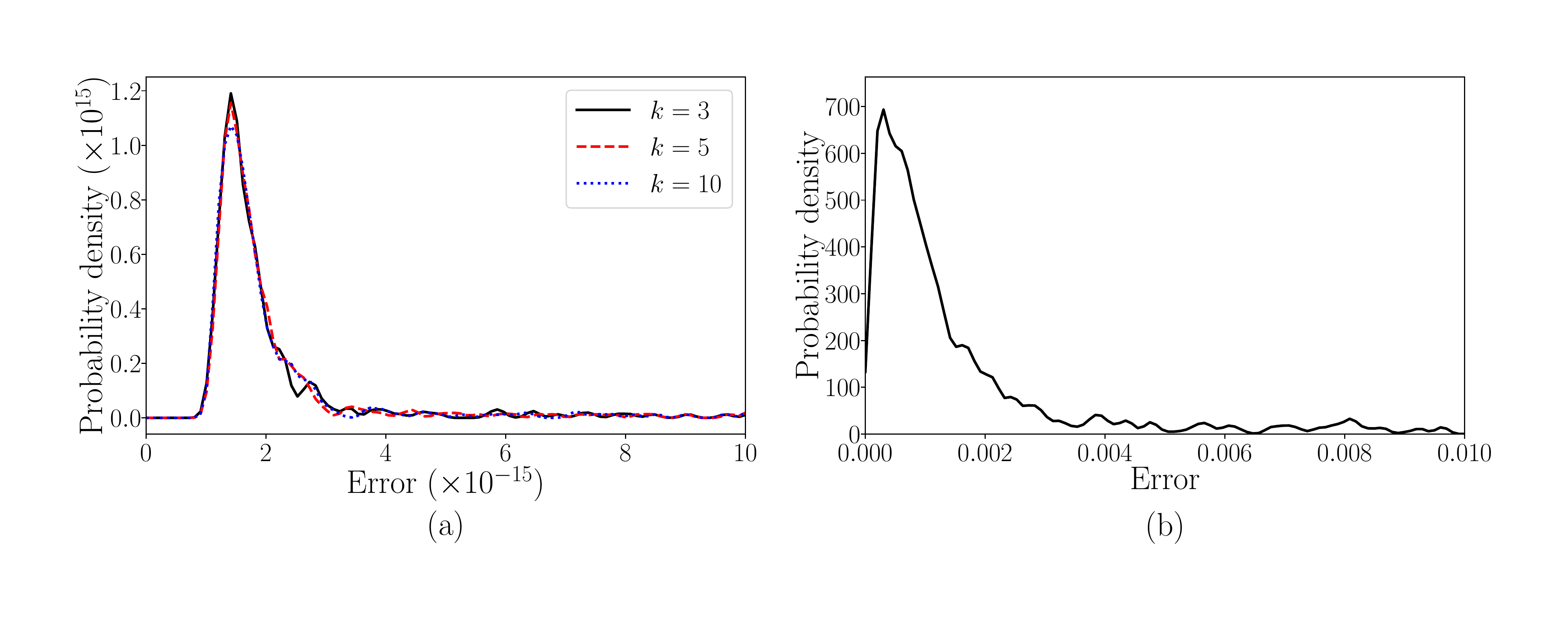}
	\caption{Example 2: PDFs of the errors for: a) local maps for different $k$ (300 samples), b) overall response prediction using all maps from 1,000 additional samples.}
	\label{fig:ex2_pdf_t}
\end{figure}

Next, we investigate the accuracy of the local maps, $\Lambda_0(\cdot)$ and $\Lambda_1(\cdot)$, between $\Delta$ (Grassmannian diffusion space) and the tangent spaces $\mathcal{T}_{\hat{\mu}_u}$ and $\mathcal{T}_{\hat{\mu}_v}$, respectively. The exact diffusion coordinates for each of the 300 training points are used, and the prediction of the corresponding points on the tangent spaces are obtained using the local maps with different numbers of neighbors ($k$) used to construct the local geometric harmonics maps. The probability density functions (PDFs) for the $\mathrm{error}_{L_2}$ are estimated using kernel density estimation for $k=$3, 5, and 10 (closest neighbors points) in Figure \ref{fig:ex2_pdf_t}a. We clearly see that the errors induced by these local maps are minimal and that they are not strongly influenced by $k$ in this specific problem, because a point is predicted in a region close to the reference point where the tangent space is constructed on. Note also that the length-scale parameter for the local maps is taken to be 0.25 times the square value of the median of the pairwise distances of the $k$ neighbors. 

The cumulative error associated with the full process is analyzed by drawing 1,000 sample points from the parameter space and computing posterior error estimation ($\mathrm{error}_{L_2}$) of the predicted response. The estimated PDF of the error is shown in Fig. \ref{fig:ex2_pdf_t}b, where the mean is equal to $1.8489 \times 10^{-3}$ and the standard deviation is equal to $2.7598 \times 10^{-3}$. This reveals that the overall errors in the prediction solutions are very small compared to their true solutions, even considering a training set of only 300 points.

\begin{figure}[!ht]
	\centering
	\captionsetup{justification=centering}
	\includegraphics[width=1\textwidth]{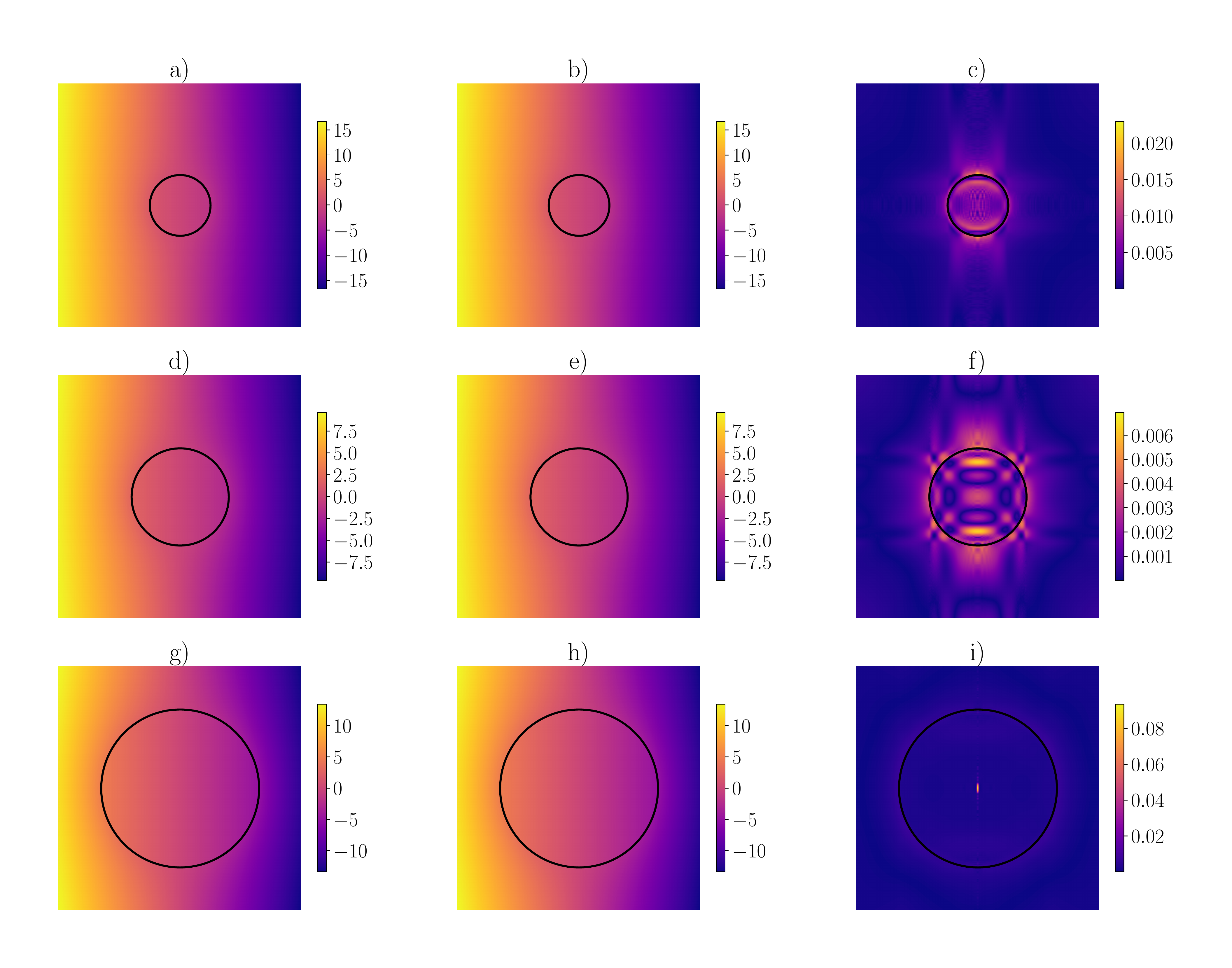}
	\vspace{-1.5em}
	\caption{Example 2: Left -- Exact electric potential for a) $\mathbf{\Theta}^{*}_a$, d) $\mathbf{\Theta}^{*}_b$, and g) $\mathbf{\Theta}^{*}_c$. Center -- Predicted electric potential for b) $\mathbf{\Theta}^{*}_a$, e) $\mathbf{\Theta}^{*}_b$, and h) $\mathbf{\Theta}^{*}_c$. Right -- Relative errors for c) $\mathbf{\Theta}^{*}_a$, f) $\mathbf{\Theta}^{*}_b$, and i) $\mathbf{\Theta}^{*}_c$}
	\vspace{-0.5em}
	\label{fig:ex2_compara_abc}
\end{figure}

Finally, for illustration we consider three test points $\mathbf{\Theta}^{*}_a = (r_0, E_{\infty}) = (0.25, 17)$, $\mathbf{\Theta}^{*}_b = (r_0, E_{\infty}) = (0.40, 10)$ and $\mathbf{\Theta}^{*}_c = (r_0, E_{\infty}) = (0.65,15)$. The exact electrical potential fields are obtained by numerically solving the model in Eq. \ref{eq:poisson_eq_gen} (Figs. \ref{fig:ex2_compara_abc}a,d,g) on a $300\times 300$ meshed domain. The predict electric fields (Figs. \ref{fig:ex2_compara_abc}b,e,h) are obtained using the surrogate model developed herein. The exact and predicted solutions are compared by $\mathrm{error}_{rel}$ in Figs. \ref{fig:ex2_compara_abc}c,f,i, where the corresponding error in the $L_2$-norm sense ($\mathrm{error}_{L_2}$) are $5.4862 \times 10^{-4}, 4.3704 \times 10^{-4}$, and $1.8306 \times 10^{-3}$ for $\mathbf{\Theta}^{*}_a$, $\mathbf{\Theta}^{*}_b$, and $\mathbf{\Theta}^{*}_c$, respectively. We can see that our surrogate model, which reduces the dimension of the solution from 90,000 spatial points to a single Grassmannian diffusion coordinate is very accurate.

\subsection{Continuum modeling of plasticity in an amorphous solid}
\label{ex:3}
An important theoretical hypothesis about the behavior of amorphous solids (e.g., metallic glasses) says that irreversible plastic deformation is mediated through atomic rearrangements in small clusters of atoms known as shear transformation zones (STZs) \cite{falk1998}. Consequently, amorphous materials subject to large shear stresses are often prone to the formation of shear bands due to the rearrangements of STZs in localized regions. It has been proposed that one can connect these large-scale plastic deformations to an effective temperature \cite{bouchbinder2009nonequilibrium} 
\begin{equation}\label{eq:teff}
    T_{eff} = \frac{\partial U_c}{\partial S_c}
\end{equation}
where $U_c$ and $S_c$ are the potential energy and entropy of the configurational degrees of freedom under the assumption that both the total energy $U$ and total entropy $S$ are separated into kinetic and configurational components, i.e. $U = U_c + U_k$ and $S = S_c + S_k$, respectively. This effective temperature provides a measure of the degree of structural disorder (characterizing the density of STZs) and can be dedimensionalized as
\begin{equation}\label{eq:teff_un}
    \chi = \frac{k_{B}T_{eff}}{E_z}
\end{equation}
where $k_B = 1.38 \times 10^{-23}$ is the Boltzman factor, $E_z$ is the STZ formation energy.

Given a spatially varying initial effective temperature field $\chi$ on a material domain, the STZ theory defines two coupled equations to model the evolution of plastic strain in the material. The first describes a plastic flow rule that relates the plastic rate of deformation tensor $\mathbf{D}^{pl}$ to the effective temperature as:
\begin{equation}\label{eq:D_pl}
    \mathbf{D}^{pl} = \frac{1}{\tau_0}\mathrm{exp}\left\{    -\left( \frac{e_z}{k_B\chi} + \frac{\Delta_{\star}}{k_B T} \right)   \right\}\mathrm{cosh}\left(  \frac{\Omega \epsilon_0 \overline{\sigma}}{k_B T}  \right)\left(1 - \frac{\sigma_y}{\overline{\sigma}}\right)
\end{equation}
Note that this flow rule is monotonic with respect to $\overline{\sigma}/\sigma_y$, with $\overline{\sigma} = |\boldsymbol{\sigma_0}|$ given as the magnitude of the deviatoric shear stress $\boldsymbol{\sigma_0} = \boldsymbol{\sigma} = \frac{1}{3}\boldsymbol{1}\mathrm{Tr}(\boldsymbol{\sigma})$. Therefore, plastic deformation does not occur when $\overline{\sigma}/\sigma_y<1$. 

The second equation describes the evolution of $\chi$ as
\begin{equation}\label{eq:chi_pl}
    c_0\dot{\chi} = \frac{1}{\sigma_y}\left( \mathbf{D}^{pl}:\boldsymbol{\sigma}_0\right) \left(\chi_{\infty} - \chi \right) + \nabla \cdot D_{\chi}\nabla\chi
\end{equation}
where $D_{\chi} = l^2\sqrt{\mathbf{D}^{pl}:\mathbf{D}^{pl}}$. Other parameters are defined in Table \ref{tab:stz}. 
\begin{center}
\begin{table}[!ht]
\centering
\caption{Parameters for the STZ plasticity model for a bulk metallic glass material.}
\begin{tabular}{cccc}
\toprule
\textbf{Parameter} & \textbf{Unit} & \textbf{Value} & \textbf{Description} \\ \toprule
 $\sigma_y$      & GPa & 0.7 & Yield stress \\
 $\tau_0$        & s & $10^{-13}$ & Molecular vibration timescale \\
 $\varepsilon_0$    & - & 0.333 & Typical local strain at STZ transition \\
 $\Delta_{\star}/k_B$    & K & 7948 & Typical activation temperature \\
 $\Omega/k_B$    & $\mathrm{\AA}^3$ & 349 & Typical activation volume \\
 $T$             & K & 97 & Bath temperature \\
 $\chi_{\infty}$ & K & 1050.6 & Steady-state effective temperature \\
 $e_z/k_B$       & K & 21000 & STZ formation energy \\
 $c_0$           & - & 0.414 & Plastic work fraction \\
 $l_{\chi}$      & $\mathrm{\AA}$ & 10 & Diffusion length scale \\ \bottomrule
\end{tabular}
\label{tab:stz}
\end{table}\label{tab:1}
\end{center}

In this example, a numerical scheme developed by Rycroft et al.\cite{rycroft2008,boffi2020} is used to solve the system of Eqs. (\ref{eq:D_pl}-\ref{eq:chi_pl}) . This method utilizes an Eulerian finite-difference method under quasi-static conditions. As mentioned previously, the STZ theory assumes that the effective temperature has a spatial distribution that influences the material response. Therefore, the evolution of Eqs. (\ref{eq:D_pl}-\ref{eq:chi_pl}) depends on the initial $\chi$ field. This field is assumed to be Gaussian \cite{hinkle2017,giovanis2020,kontolati2021manifold}, therefore it can be characterized by the mean $\mu_{\chi}$ and coefficient of variation $c_{\chi}$. Herein, it is assumed that these parameters are both uncertain with uniform distributions as described in Table \ref{tab:rv_stz}.  The associated correlation structure and length-scale are inferred from molecular dynamics simulations~\cite{hinkle2017,konakli2016}. In the simulations, a simple shear up to 50\% strain is imposed to a simulation box of size 400\AA × 400\AA. A grid of size $32 \times 32$ is considered in the discretization, where each element has a size of 12.5\AA $\times$ 12.5\AA. Therefore, each snapshot of this simulation is given by a matrix $\mathbf{X}_i \in \mathbb{R}^{32 \times 32}$.
\begin{center}
\begin{table}[!ht]
\centering
\caption{Probability distributions of the STZ random field parameters.}
\vspace{-7pt}
\begin{tabular}{l l c c}
\toprule
Description of variables/parameters & \hspace{15pt} &  & \hspace{2pt} Uncertainty/value \\ [0.5ex]
\toprule
Mean  &   & $\mu_{\mathcal{X}}$ & $\sim \mathcal{U}(500, 700)$\vspace{2pt}  \\ 
Coefficient of variation   &   &   $c_{\mathcal{X}}$ & $\sim \mathcal{U}(0, 0.1)$  \vspace{1pt} \\ 
\bottomrule
\end{tabular}
\label{tab:rv_stz}
\end{table}
\end{center}

\begin{figure}[!ht]
	\centering
	\captionsetup{justification=centering}
	\includegraphics[scale=0.15]{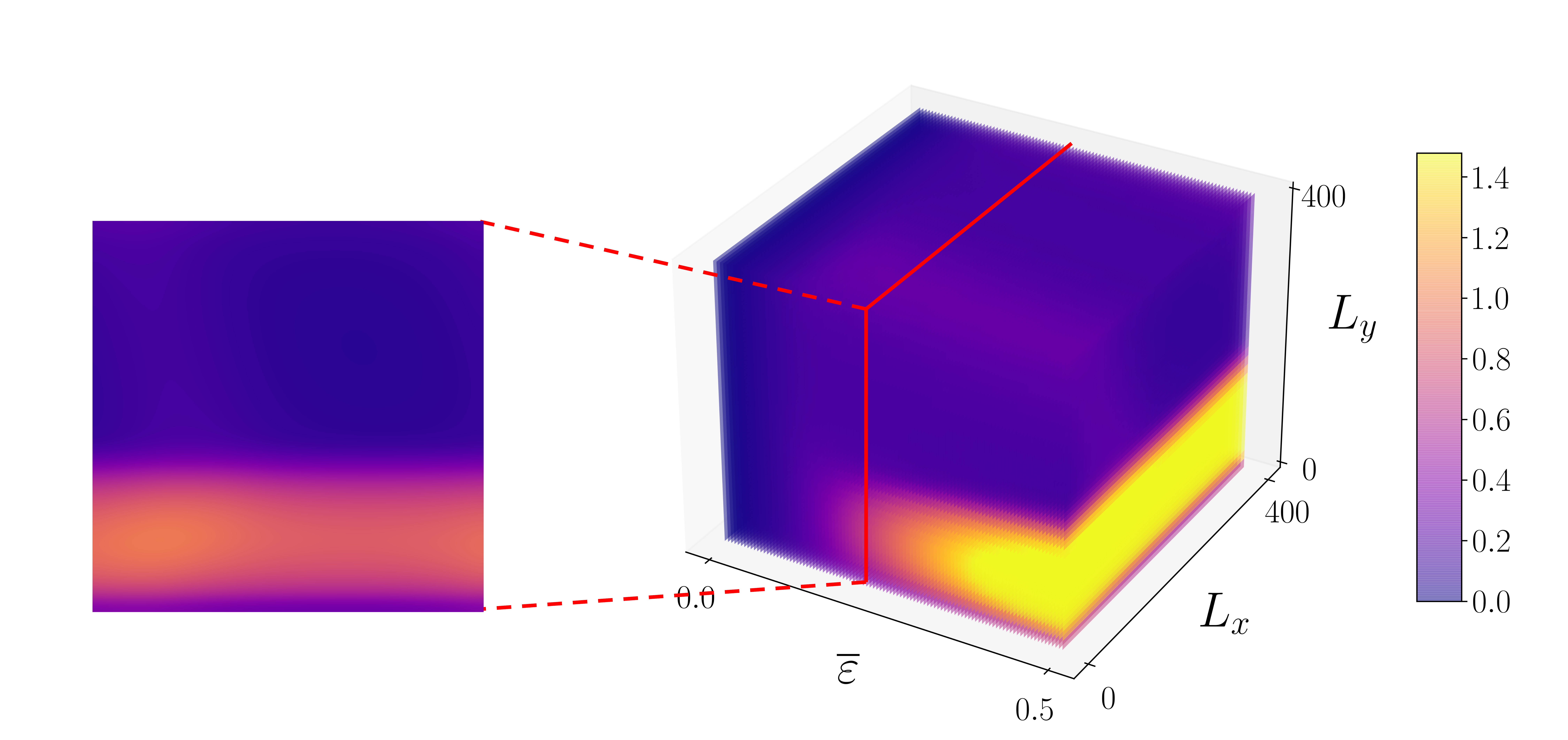}
	\vspace{-1.0em}
	\caption{Evolution of plastic strain with a snapshot of the strain field at a given strain level extracted.}
	\vspace{-0.5em}
	\label{fig:ex3_slice}
\end{figure}

We obtain $N = 196$ samples of the pair $\mathbf{\Theta}=(\mu_{\chi}, c_{\chi})$ via stratified sampling to train a surrogate model for full evolution of the plastic strain field. The evolution of the plastic strain field for a given pair $\mathbf{\Theta}=(\mu_{\chi}, c_{\chi})$ is presented in Fig. \ref{fig:ex3_slice} as a sequence of 101 snapshots of size $L_x$ and $L_y$ at discrete values of the imposed shear strain $\bar{\epsilon}$. A matrix $\boldsymbol{Y} \in \mathbb{R}^{1024 \times 101}$ is then constructed for a given pair $\mathbf{\Theta}=(\mu_{\chi}, c_{\chi})$, where each column of $\boldsymbol{Y}$ correspond to the vectorized snapshot of the plastic strain field.

\begin{figure}[!ht]
	\centering
	\captionsetup{justification=centering}
	\includegraphics[scale=0.3]{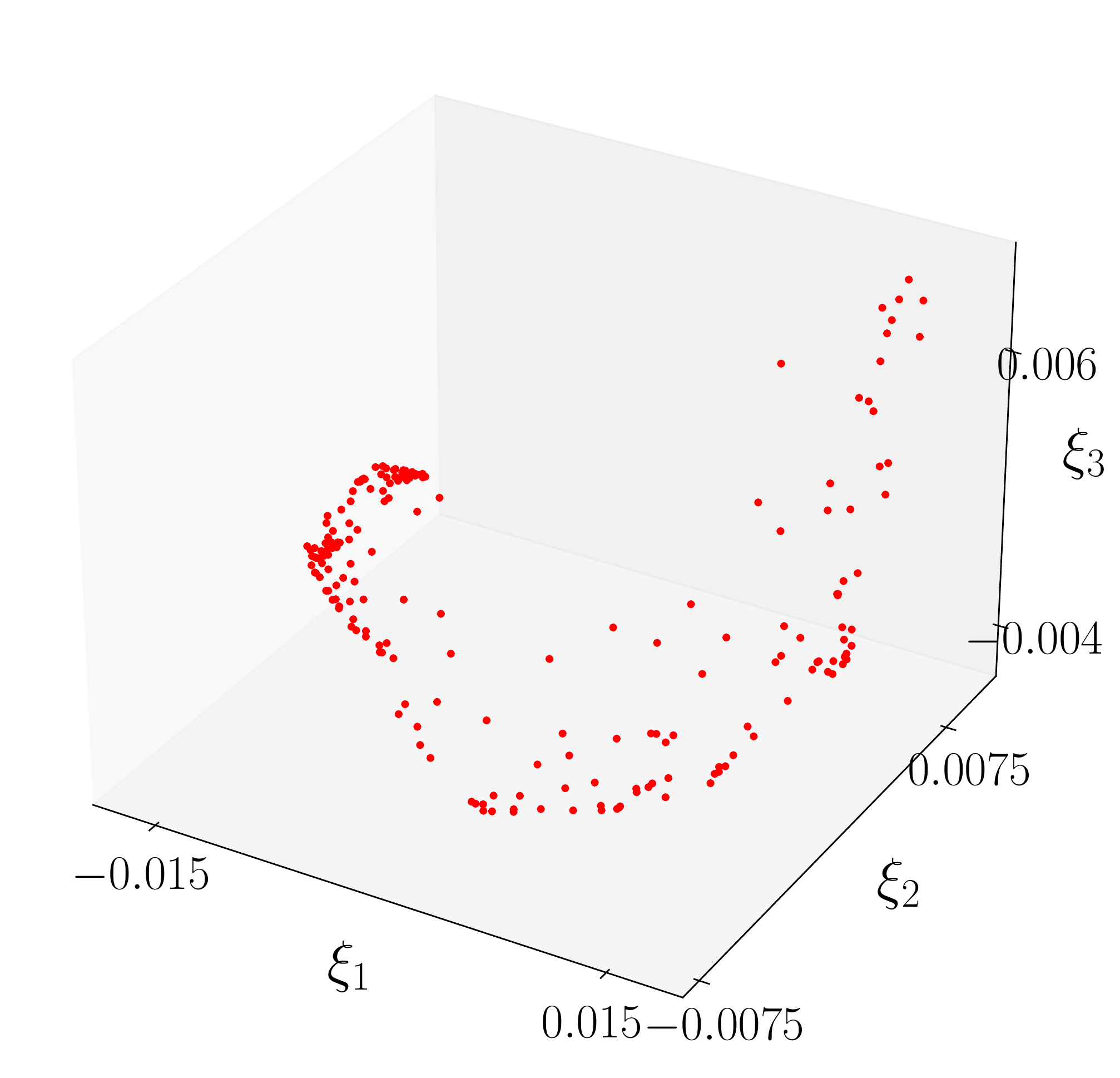}
	\vspace{-0.5em}
	\caption{Grassmannian diffusion coordinates.}
	\vspace{-0.5em}
	\label{fig:ex3_gdmaps}
\end{figure}

In this problem, two Grassmann manifolds given by $\mathcal{G}(p, n)$ (left manifold) and $\mathcal{G}(p, m)$ (right manifold) are associated with the left and right singular vectors of the matrices $\boldsymbol{Y}$; where $n=1024$, $m=101$, and $p=10$ suffices to encode the geometric structure of each data point. Using GDMaps, we obtain a set of 196 Grassmannian diffusion coordinates embedding the high-dimensional data into a low-dimensional Euclidean space as shown in Fig. \ref{fig:ex3_gdmaps} with $r=3$. 

\begin{figure}[H]
	\centering
	\captionsetup{justification=centering}
	\includegraphics[width=0.7\textwidth]{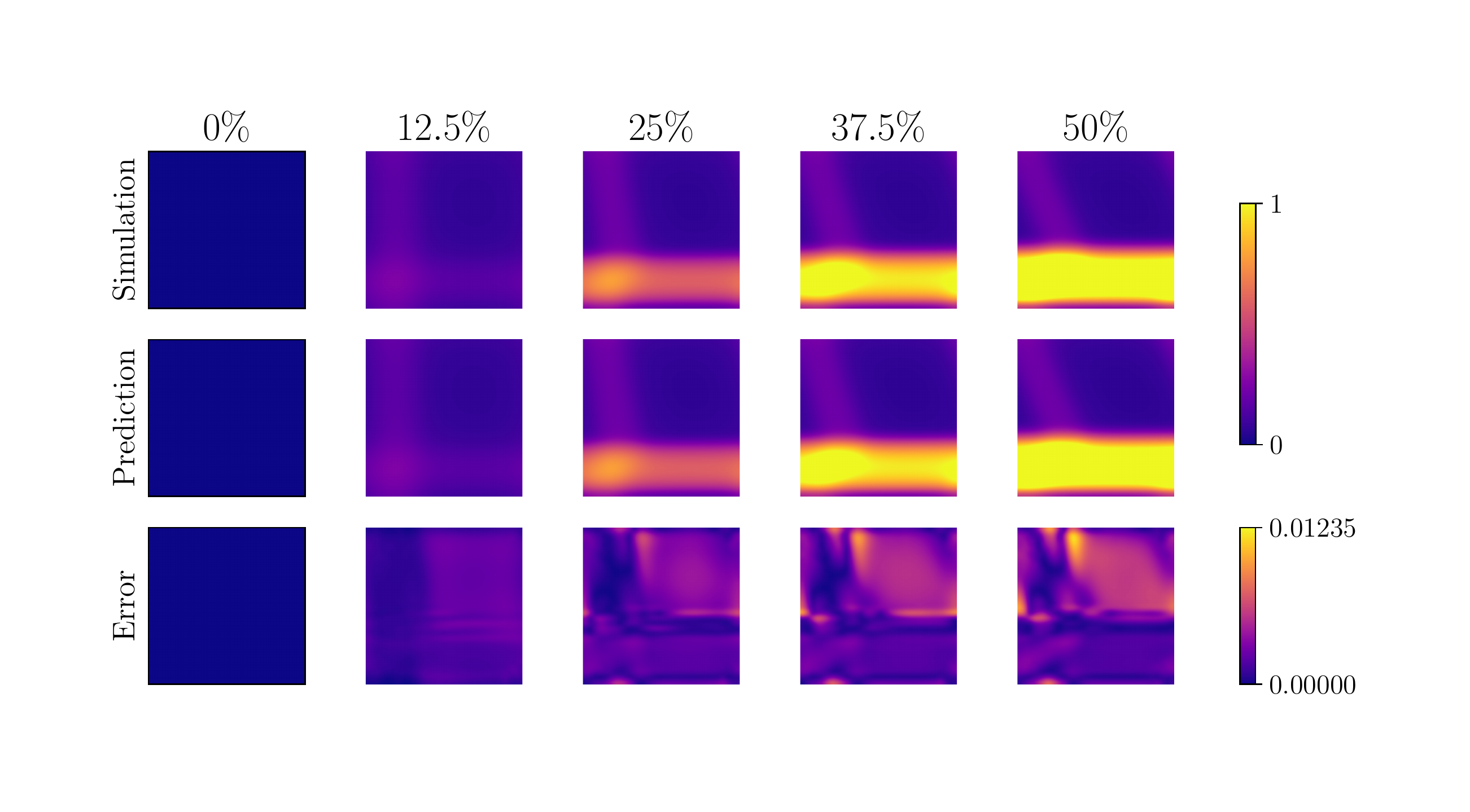}
	\vspace{-0.5em}
	\caption{Exact and simulated strain field evolution for $(\mu_{\chi}, c_{\chi}) = (530.1748, 0.079)$ and the corresponding errors.}
	\vspace{-0.5em}
	\label{fig:ex3_comp}
\end{figure}

\begin{figure}[H]
	\centering
	\captionsetup{justification=centering}
	\includegraphics[width=0.7\textwidth]{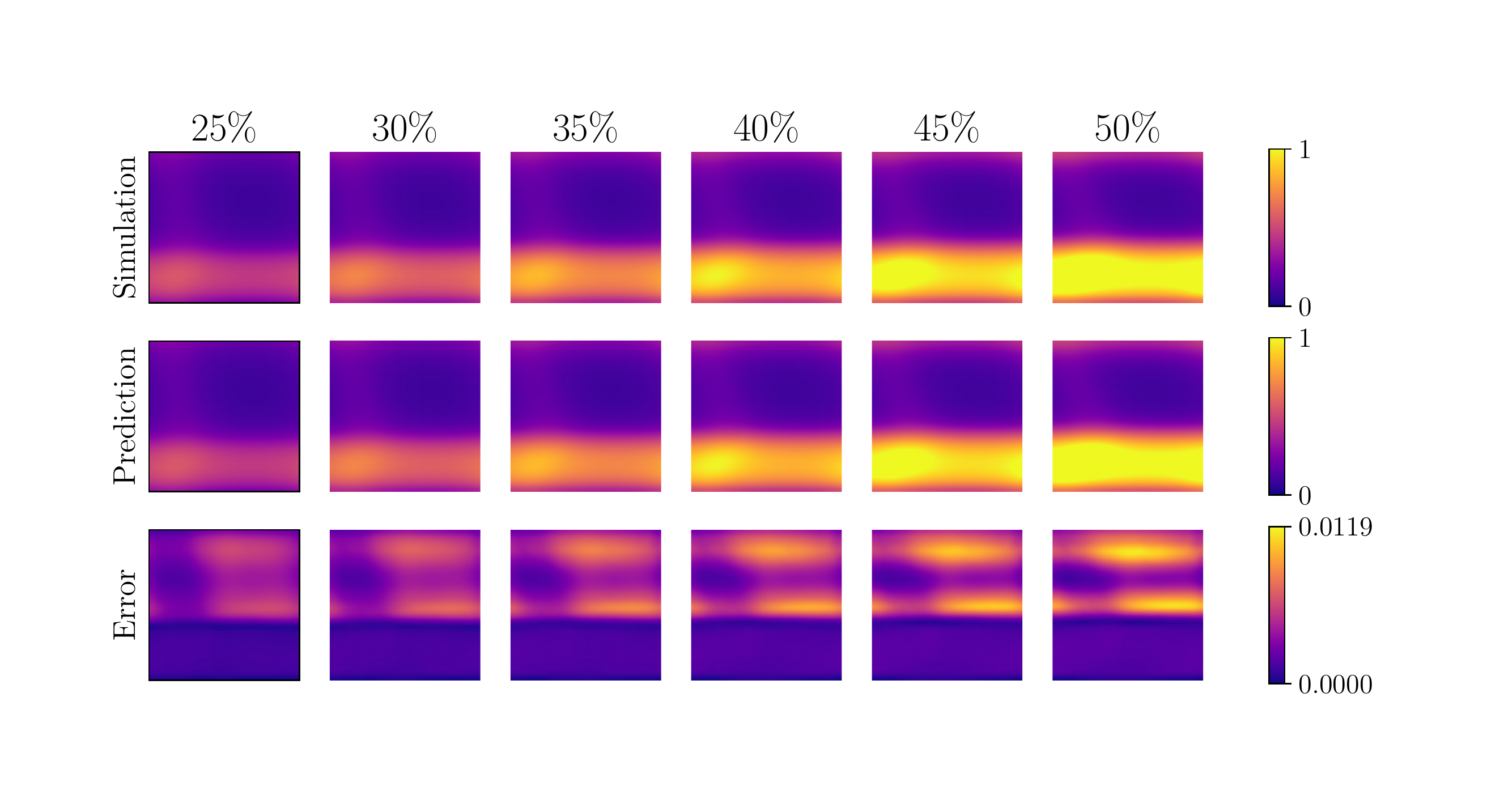}
	\vspace{-0.5em}
	\caption{Simulated and predicted evolution of the mean of the strain field for 100 additional samples and the corresponding errors.}
	\vspace{-0.5em}
	\label{fig:ex3_comp_mean}
\end{figure}

\begin{figure}[H]
	\centering
	\captionsetup{justification=centering}
	\includegraphics[width=0.7\textwidth]{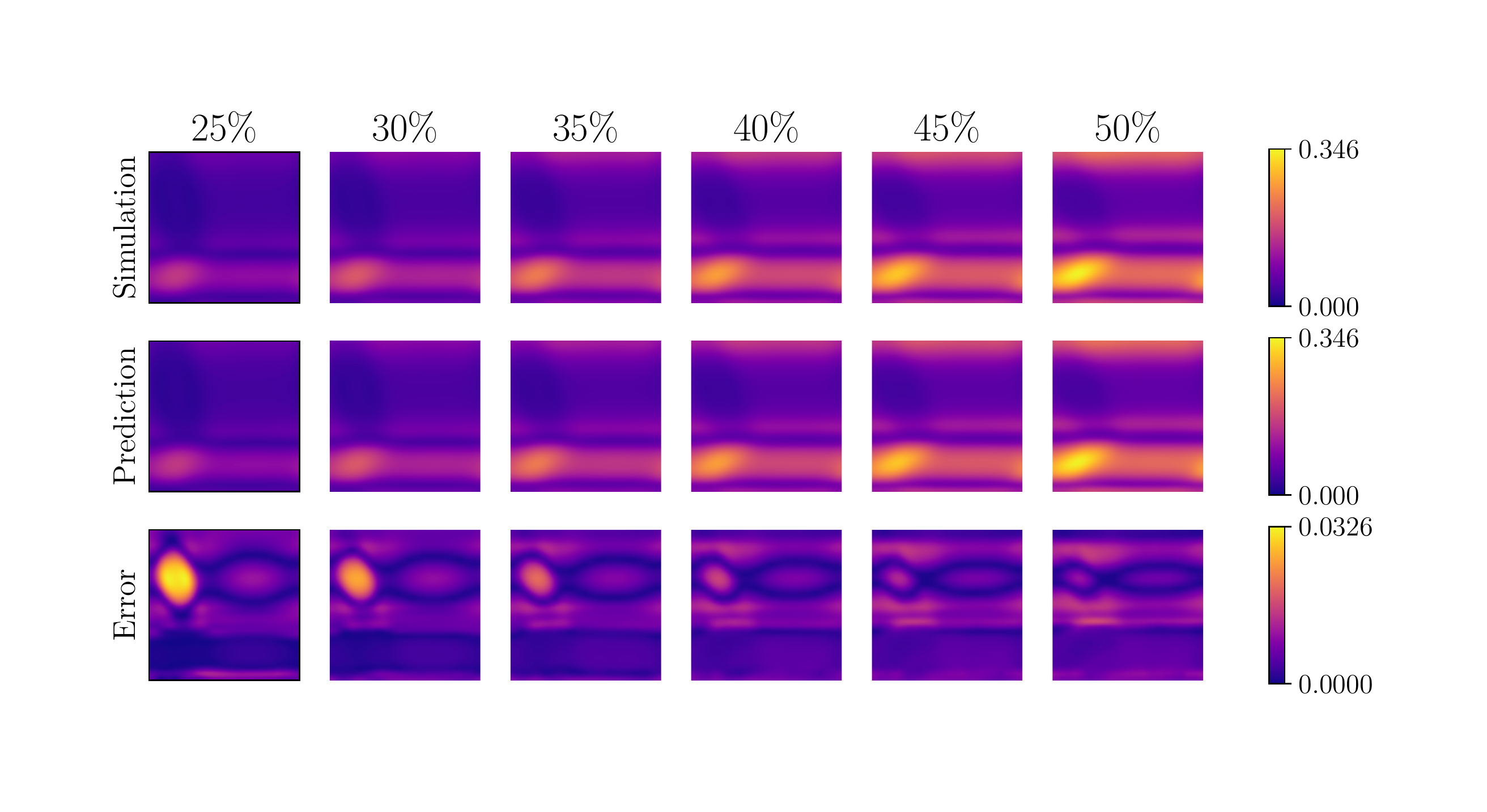}
	\vspace{-0.5em}
	\caption{Simulated and predicted evolution of the standard deviation of the strain field for 100 additional samples and the corresponding errors.}
	\vspace{-0.5em}
	\label{fig:ex3_comp_std}
\end{figure}

Once the surrogate model is trained using the 196 samples obtained previously, we can predict the full evolution of the plastic strain field for any pair $\mathbf{\Theta}=(\mu_{\chi}, c_{\chi})$. Considering a representative case with $\mathbf{\Theta}= (530.1748, 0.0792)$, the simulated and predicted evolution of the plastic strain field, as well as their relative error, are presented in Fig. \ref{fig:ex3_comp} for five different levels of imposed strain ($0\%, 12.5\%, 25\%, 37.5\%, 50\%$). The error in the $L_2$-norm sense for this plastic strain field is equal to $2.0473 \times 10^{-3}$. Next, considering 100 additional samples we compute the mean and standard deviation of the plastic strain fields at the different levels of imposed strain ($25\%, 30\%, 35\%, 40\%, 45\%, 50\%$) as presented in Figs. \ref{fig:ex3_comp_mean}, and \ref{fig:ex3_comp_std}. These figures include the statistical characterization obtained by using the numerical model and the surrogate model, along with the relative errors. From these results, we see that the surrogate model developed herein can predict the uncertain response of a complex model with high-accuracy, by taking advantage of low-dimensional subspace structure of the problem to reduce the computational burden associated with running high-fidelity models for UQ.

\section{Conclusions}\label{sec5}

This paper introduces a fully data-driven surrogate model for uncertainty quantification of high-dimensional models of complex physical/engineering systems. It takes advantage of the low-dimensional representation of high-dimensional input/output data obtained via Grassmannian diffusion maps to create a set of geometric harmonics based maps. A global map is constructed to predict the Grassmannian diffusion coordinates corresponding to any new element in the set of input parameters with good accuracy. Once the Grassmannian diffusion coordinates corresponding to a new set of input parameters are predicted, the $k$-nearest neighbors points in the Grassmannian diffusion space and their associated points on the Grassmann manifold are utilized to estimate, via geometric harmonics, a local map from the Grassmannian diffusion space to a tangent space. Next, the exponential map project the point onto the Grassmann manifold, a result used to predict the high-fidelity solution of the problem.

The method developed herein used the descriptive power of the Grassmannian diffusion maps and the computational performance of geometric harmonics to provide an efficient and accurate prediction of the solution of complex systems described by algebraic equations and partial/ordinary differential equations. Three examples were considered to evaluate the performance of this technique. The first one consisted of a toy example to demonstrate the ability of the technique to predict data with complex geometry using spectral methods in a way that is easy to understand and visualize. In the second example, the performance of the surrogate modeling developed herein was verified in a physical model (i.e., electric potential of a cylinder in homogeneous electric field) with high-dimensional response, also considering discontinuities in the system response. It was demonstrated that some parameters such as the length-scale parameter and the number of retained eigenvalues, both for the Gaussian kernel used in the geometric harmonics framework; as well as the amount of data in the training set, are important quantities affecting the accuracy of the presented technique. The third problem analyzed in this paper evaluated the plastic deformation of amorphous solids using the shear transformation zone (STZ) theory of plasticity. The uncertainty was imposed in the mean and coefficient of variation of the initial nondimensionalized effective temperature field ($\chi$). In this case, the evolution of the strain field with the strain level is also taken into consideration, and the uncertainties of the plastic strain field are predicted accurately. 

In all the cases considered herein, a good accuracy was identified in the predicted solutions in comparison with the exact ones. The method proves advantageous due to its computational performance and ability to make reliable predictions for high-dimensional responses considering a highly sparse set of points.

\section*{Acknowledgments}

This material is based upon work supported by the U.S. Department of Energy, Office of Science, Office of Advanced Scientific Computing Research under Award Number DE-SC0020428.

\subsection*{Financial disclosure}

None reported.

\subsection*{Conflict of interest}

The authors declare no potential conflict of interests.

\bibliographystyle{siamplain}
\bibliography{mybibfile}%

\end{document}